\documentclass{article}
\usepackage{amsfonts}
\usepackage{amssymb}
\usepackage{amsmath}

\input{tcilatex}
\begin{document}

\title{The Entropic Dynamics approach to Quantum Mechanics}
\author{Ariel Caticha}
\date{ }
\maketitle

\begin{abstract}
Entropic Dynamics (ED) is a framework in which Quantum Mechanics is derived
as an application of entropic methods of inference. In ED the dynamics of
the probability distribution is driven by entropy subject to constraints
that are codified into a quantity later identified as the phase of the wave
function. The central challenge is to specify how those constraints are
themselves updated. In this paper we review and extend the ED framework in
several directions. A new version of ED is introduced in which particles
follow smooth differentiable Brownian trajectories (as opposed to
non-differentiable Brownian paths). To construct ED we make use of the fact
that the space of probabilities and phases has a natural symplectic
structure (i.e., it is a phase space with Hamiltonian flows and Poisson
brackets). Then, using an argument based on information geometry, a metric
structure is introduced. It is shown that the ED that preserves the
symplectic and metric structures -- which is a Hamilton-Killing flow in
phase space -- is the linear Schr\"{o}dinger equation. These developments
allow us to discuss why wave functions are complex and the connections
between the superposition principle, the single-valuedness of wave
functions, and the quantization of electric charges. Finally, it is observed
that Hilbert spaces are not necessary ingredients in this construction. They
are a clever but merely optional trick that turns out to be convenient for
practical calculations.
\end{abstract}

\section{Introduction}

Quantum mechanics has been commonly regarded as a generalization of
classical mechanics with an added element of indeterminism. The standard
quantization recipe starts with a description in terms of the system's
classical coordinates and momenta $\{q,p\}$ and then proceeds by applying a
series of more or less ad hoc rules that replace the classical $\{q,p\}$ by
self-adjoint linear operators $\{\hat{q},\hat{p}\}$ acting on some complex
Hilbert space \cite{Dirac 1930}. The Hilbert space structure is given
priority while the probabilistic structure is relegated to the less
fundamental status of providing phenomenological rules for how to handle
those mysterious physical processes called measurements. The result is a
dichotomy between two separate and irreconcilable modes of wave function
evolution: one is the linear and deterministic Schr\"{o}dinger evolution and
the other is the discontinuous and stochastic wave function collapse \cite%
{von Neumann 1955}\cite{Bell 1990}. To put it bluntly, the dynamical and the
probabilistic aspects of quantum theory are incompatible with each other.
And furthermore, the dichotomy spreads to the interpretation of the quantum
state itself.\footnote{%
Excellent reviews with extended references to the literature are given in 
\emph{e.g.} \cite{Stapp 1972}-\cite{Leifer 2014}.} It obscures the issue of
whether the wave function describes the \emph{ontic} state of the system or
whether it describes an \emph{epistemic} state about the system.\footnote{%
Since the terms `ontic' and `epistemic' are not yet of widespread use
outside the community of Foundations of QM a clarification might be useful.
A concept is referred as `ontic' when it describes something that is
supposed to be real, to exist out there independently of any observer. A
concept is referred as `epistemic' when it is related to the state of
knowledge, opinion, or belief of an agent, albeit an ideally rational agent.
Examples of epistemic quantities are probabilities and entropies. An
important point is that the distinction ontic/epistemic is not the same as
the distinction objective/subjective. For example, probabilities are fully
epistemic --- they are tools for reasoning with incomplete information ---
but they can lie anywhere in the spectrum from being completely subjective
(two different agents can have different beliefs) to being completely
objective. In QM, for example, probabilities are epistemic and objective.
Indeed, at the non-relativistic level anyone who computes probabilities that
disagree with QM will be led to experimental predictions that are
demonstrably wrong. We will say that the wave function $\Psi $, which is
fully epistemic and objective, represents a \textquotedblleft
physical\textquotedblright\ state when it represents information about an
actual \textquotedblleft physical\textquotedblright\ situation.}

In the Entropic Dynamics (ED) approach these problems are resolved by
placing the probabilistic aspects of QM at the forefront while the Hilbert
space structure is relegated to the secondary role of a convenient
calculational tool \cite{Caticha 2010}-\cite{Caticha 2017}. ED tackles QM as
an example of entropic inference, a framework designed to handle
insufficient information.\footnote{%
The principle of maximum entropy as a method for inference can be traced to
the pioneering work of E. T. Jaynes \cite{Jaynes 1957}-\cite{Jaynes 2003}.
For a pedagogical overview including more modern developments see \cite%
{Caticha 2012}\cite{Caticha 2014}.} The starting point is to specify the
subject matter, the ontology --- are we talking about the positions of
particles or the configurations of fields? Once this decision is made our
inferences about these variables are driven by entropy subject to
information expressed by constraints. The main effort is directed towards
choosing those constraints since it is through them that the
\textquotedblleft physics\textquotedblright\ is introduced.

From the ED perspective many of the questions that seemed so urgent in other
approaches are successfully evaded. For example, when quantum theory is
regarded as an extension of classical mechanics any deviations from
causality demand an explanation. In contrast, in the entropic approach
uncertainty and probabilities are the norm. Indeterminism is just the
inevitable consequence of incomplete information and no deeper explanation
is needed. Instead, it is the certainty and determinism of the classical
limit that require explanations. Another example of a question that has
consumed an enormous effort is the problem of deriving the Born rule from a
fundamental Hilbert space structure. In the ED approach this question does
not arise and the burden of explanation runs in the opposite direction: how
do objects such as wave functions involving complex numbers emerge in a
purely probabilistic framework? Yet a third example concerns the
interpretation of the wave function itself. ED offers an uncompromising and
radically epistemic view of the wave function $\Psi $. This turns out to be
extremely restrictive: in a fully epistemic interpretation there is no
logical room for \textquotedblleft quantum\textquotedblright\ probabilities
obeying alternative rules of inference. Not only is the probability $|\Psi
|^{2}$ interpreted as a state of knowledge but, in addition, the epistemic
significance of the phase of the wave function must be clarified and made
explicit. Furthermore, it is also required that all \emph{updates of }$\Psi $%
, which include \emph{both} its unitary time evolution \emph{and} the wave
function collapse during measurement, must be obtained as a consequence of
entropic and Bayesian updating rules.\footnote{%
There exist many different Bayesian interpretations of probability. In
section \ref{final remarks} we comment on how ED differs from the frameworks
known as Quantum Bayesianism \cite{Brun et al 2001}-\cite{Caves et al 2002b}
and its closely related descendant QBism \cite{Fuchs Schack 2013}\cite{Fuchs
et al 2014}.}

There is a large literature on reconstructions of quantum mechanics (see 
\emph{e.g.}, \cite{Nelson 1985}-\cite{tHooft 2016} and references therein)
and there are several approaches based on information theory (see \emph{e.g.}%
, \cite{Wootters 1981}-\cite{DAriano 2017}). What distinguishes ED is a
strict adherence to Bayesian and entropic methods and also a central concern
with the nature of time. The issue here is that any discussion of dynamics
must inevitably include a notion of time but the rules for inference do not
mention time --- they are totally atemporal. One can make inferences about
the past just as well as about the present or the future. This means that
any model of dynamics based on inference must also include assumptions about
time, and those assumptions must be explicitly stated. In ED
\textquotedblleft entropic\textquotedblright\ time is a book-keeping device 
\emph{designed} to keep track of changes. The construction of entropic time
involves several ingredients. One must introduce the notion of an `instant';
one must show that these instants are suitably ordered; and finally one must
define a convenient measure of the duration or interval between the
successive instants. It turns out that an arrow of time is generated
automatically and entropic time is intrinsically directional.

This paper contains a review of previous work on ED and extends the
formalism in several new directions. In \cite{Caticha 2010}-\cite{Caticha
2017} the Schr\"{o}dinger equation was derived as a peculiar non-dissipative
diffusion in which the particles perform an irregular Brownian\ motion that
resembles the Einstein-Smoluchowski (ES) process \cite{Nelson 1967}. The
trajectories are continuous and non-differentiable so their velocity is
undefined. Since the expected length of the path between any two points is
infinite this would be a very peculiar motion indeed. Here we exhibit a new
form of ED in which the Brownian motion resembles the much smoother
Oernstein-Uhlenbeck (OU) process \cite{Nelson 1967}. The trajectories have
finite expected lengths; they are continuous and differentiable. On the
other hand, although the velocities are well defined and continuous, they
are not differentiable.\footnote{%
In both the ES and the OU processes, which were originally meant to model
the actual physical Brownian motion, friction and dissipation play essential
roles. In contrast, ED is non-dissipative. ED formally resembles Nelson's
stochastic mechanics \cite{Nelson 1985} but the conceptual differences are
significant. Nelson's mechanics attempted an ontic interpretation of QM as
an ES process driven by real stochastic classical forces while ED is a
purely epistemic model that does not appeal to an underlying classical
mechanics.}

We had also shown that the irregular Brownian motion at the
\textquotedblleft microscopic\textquotedblright\ or sub-quantum level was
not unique. One can enhance or suppress the fluctuations while still
obtaining the same emergent Schr\"{o}dinger behavior at the
\textquotedblleft macroscopic\textquotedblright\ or quantum level \cite%
{Bartolomeo Caticha 2015}\cite{Bartolomeo Caticha 2016}. A similar
phenomenon is also found in the smoother ED developed here. In both the ES
and the OU cases the special limiting case in which fluctuations are totally
suppressed turns out to be of particular interest because the particles
evolve deterministically along the smooth lines of probability flow. This
means that ED includes the Bohmian or causal form of quantum mechanics \cite%
{Bohm 1952}-\cite{Holland 1993} as a limiting case.

ED consists in the entropic updating of probabilities through information
supplied by constraints. The main concern is how these constraints are
chosen including, in particular, how the constraints themselves are updated.
In \cite{Caticha et al 2014} an effective criterion was found by adapting
Nelson's seminal insight that QM is a non-dissipative diffusion \cite{Nelson
1979}. This amounts to updating constraints in such a way that a certain
energy functional is conserved. Unfortunately this criterion, while fully
satisfactory in a non-relativistic setting, fails in curved space-times
where the concept of a globally conserved energy may not exist.

The second contribution in this paper is a geometric framework for updating
constraints that does not rely on the notion of a conserved energy. Our
framework draws inspiration from two sources: one is the fact that QM has a
rich geometrical structure \cite{Kibble 1979}-\cite{Elze 2012}. The authors
of \cite{Kibble 1979}-\cite{Ashtekar Schilling 1998} faced the task of
unveiling geometric structures that although well hidden are already present
in the standard QM framework. Our goal runs in the opposite direction: we
impose these natural geometric structures as the foundation upon which we
reconstruct the QM formalism.

The other source of inspiration is the connection between QM\ and
information geometry \cite{Caticha 2012}, \cite{Amari 1985}-\cite{Ay et al
2017} that was originally suggested in the work of Wootters \cite{Wootters
1981}. This connection has been explored in the context of quantum
statistical inference \cite{Brody Hughston 1997}, in the operational
description of quantum measurements \cite{Mehrafarin 2005}\cite{Goyal 2010},
and in the reconstruction of QM \cite{Reginatto Hall 2011}\cite{Reginatto
Hall 2012}. Our previous presentation in \cite{Caticha 2017} has been
considerably streamlined by recognizing the central importance of symmetry
principles when implemented in conjunction with concepts of information
geometry.

In ED the degrees of freedom are the probability densities $\rho (x)$ and
certain \textquotedblleft phase\textquotedblright\ fields $\Phi (x)$ that
represent the constraints that control the flow of probabilities. Thus we
are concerned not just with the \textquotedblleft
configuration\textquotedblright\ space of probabilities $\{\rho \}$ but with
the larger space of probabilities and phases $\{\rho ,\Phi \}$. The latter
has a natural symplectic structure, that is, $\{\rho ,\Phi \}$ is a phase
space. Imposing a dynamics that preserves this symplectic structure leads to
Hamiltonian flows, Poisson brackets, and so much of the canonical formalism
associated with mechanics. To single out the particular Hamiltonian flow
that reproduces QM we extend the information geometry of the configuration\
space $\{\rho \}$ to the full phase space. This is achieved by imposing a
symmetry that is natural in a probabilistic setting: we extend the
well-known spherically\ symmetric information geometry of the space $\{\rho
\}$ to the full phase space $\{\rho ,\Phi \}$. This construction yields a
derivation of the Fubini-Study metric. A welcome by-product is that the
joint presence of a symplectic and a metric structure leads to a complex
structure. This is the reason why QM involves complex numbers.

The dynamics that preserves the metric structure is a Killing flow. We
propose that the desired geometric criterion for updating constraints is a
dynamics that preserves both the symplectic and the metric structures. Thus,
in\ the final step of our reconstruction of QM we show that the Hamiltonians
that generate Hamiltonian-Killing flows lead to an entropic dynamics
described by the linear Schr\"{o}dinger equation.

We conclude with some comments exploring various aspects of the ED
formalism. We show that, despite the arrow of entropic time, the resulting
ED is symmetric under time reversal. We discuss the connections between
linearity, the superposition principle, the single-valuedness of wave
functions, and the quantization of charge. We also discuss the classical
limit and the Bohmian limit in which fluctuations are suppressed and
particles follow deterministic trajectories. Finally, we discuss the
introduction of Hilbert spaces. We argue that, while strictly unnecessary in
principle, Hilbert spaces are extremely convenient for calculational
purposes.

This paper focuses on the derivation of the Schr\"{o}dinger equation but the
ED approach has been applied to a variety of other topics in quantum theory.
These include: the quantum measurement problem \cite{Johnson Caticha 2011}%
\cite{Vanslette Caticha 2016}; momentum and uncertainty relations \cite%
{Nawaz Caticha 2011}\cite{Bartolomeo Caticha 2016};\footnote{%
These are the well-known uncertainty relations due to Heiseberg and to Schr%
\"{o}dinger. The entropic uncertainty relations proposed by Deutsch \cite%
{Deutsch 1983}-\cite{Maassen Uffink 1988} have not yet been explored within
the context of ED.} the Bohmian limit \cite{Bartolomeo Caticha 2015}\cite%
{Bartolomeo Caticha 2016} and the classical limit \cite{Demme Caticha 2016};
extensions to curved spaces \cite{Nawaz et al 2015}; to relativistic fields 
\cite{Ipek Caticha 2014}\cite{Ipek Abedi Caticha 2018}; and the ED of spin 
\cite{Caticha Carrara 2019}.

\section{The ED of short steps}

We deal with $N$ particles living in a flat 3-dimensional space $\mathbf{X}$
with metric $\delta _{ab}$. For $N$ particles the configuration space is $%
\mathbf{X}_{N}=\mathbf{X}\times \ldots \times \mathbf{X}$. We assume that
the particles have definite positions $x_{n}^{a}$ and it is their unknown
values that we wish to infer.\footnote{%
In this work ED is a model for the quantum mechanics of particles. The same
framework can be deployed to construct models for the quantum mechanics of
fields, in which case it is the fields that are ontic\ and have well-defined
albeit unknown values \cite{Ipek Caticha 2014}\cite{Ipek Abedi Caticha 2018}}
(The index $n$ $=1\ldots N$ denotes the particle and $a=1,2,3$ the spatial
coordinates.)

In ED positions play a very special role: they define the ontic state of the
system. This is in contradiction with the standard Copenhagen notion that
quantum particles acquire definite positions only as a result of a
measurement. For example, in the ED description of the double slit
experiment the particle definitely goes through one slit or the other but
one might not know which. The wave function, on the other hand, is a purely
epistemic notion and, as it turns out, all other quantities, such as energy
or momentum, are epistemic too. They do not reflect properties of the
particles but properties of the wave function \cite{Johnson Caticha 2011}%
\cite{Vanslette Caticha 2016}\cite{Nawaz Caticha 2011}.

Having identified the microstates $x\in \mathbf{X}_{N}$ we tackle the
dynamics. The main dynamical assumption is that the particles follow
trajectories that are continuous. This represents an enormous simplification
because it implies that a generic motion can be analyzed as the accumulation
of many infinitesimally short steps. Therefore, the first task is to find
the transition probability $P(x^{\prime }|x)$ for a short step from an
initial $x$ to an unknown neighboring $x^{\prime }$ and only later we will
determine how such short steps accumulate to yield a finite displacement.

The probability $P(x^{\prime }|x)$ is found by maximizing the entropy 
\begin{equation}
\mathcal{S}[P,Q]=-\int dx^{\prime }\,P(x^{\prime }|x)\log \frac{P(x^{\prime
}|x)}{Q(x^{\prime }|x)}~  \label{entropy a}
\end{equation}%
relative to the joint prior $Q(x^{\prime }|x)$ subject to constraints given
below. (In multidimensional integrals such as (\ref{entropy a}) the notation 
$dx^{\prime }$ stands for $d^{3N}x^{\prime }$.)

\paragraph*{The prior ---}

The choice of prior $Q(x^{\prime }|x)$ must reflect the state of knowledge
that is common to all short steps. (It is through the constraints that the
information that is specific to any particular short step will be supplied.)
We adopt a prior that carries the information that the particles take
infinitesimally short steps and also reflects the translational and
rotational invariance of the Euclidean space $\mathbf{X}$ but is otherwise
uninformative. In particular, the prior expresses total ignorance about any
correlations. Such a prior can itself be derived from the principle of
maximum entropy. Indeed, maximize 
\begin{equation}
S[Q]=-\int dx^{\prime }\,Q(x^{\prime }|x)\log \frac{Q(x^{\prime }|x)}{\mu
(x^{\prime })}~,
\end{equation}%
relative to the uniform measure $\mu (x^{\prime })$,\footnote{%
In Cartesian coordinates $\mu =\limfunc{const}$ and may be ignored.} subject
to normalization, and subject to the $N$ independent constraints 
\begin{equation}
\langle \delta _{ab}\Delta x_{n}^{a}\Delta x_{n}^{b}\rangle =\kappa
_{n}~,\quad (n=1\ldots N)~,
\end{equation}%
where $\kappa _{n}$ are small constants and $\Delta x_{n}^{a}=x_{n}^{\prime
a}-x_{n}^{a}$. The result is a product of Gaussians, 
\begin{equation}
Q(x^{\prime }|x)\propto \exp -\frac{1}{2}\dsum\nolimits_{n}\alpha _{n}\delta
_{ab}\Delta x_{n}^{a}\Delta x_{n}^{b}~,  \label{prior}
\end{equation}%
where, in order to reflect translational invariance and possibly
non-identical particles, the Lagrange multipliers $\alpha _{n}$ are
independent of $x$ but may depend on the index $n$. Eventually we will let $%
\alpha _{n}\rightarrow \infty $ in order to implement infinitesimally short
steps. Next we specify the constraints that are specific to each particular
short step.

\paragraph*{The drift potential constraint ---}

In Newtonian dynamics one does not need to explain why a particle perseveres
in its motion in a straight line; what demands an explanation --- that is, a
force --- is why the particle deviates from inertial motion. In ED one does
not require an explanation for why the particles move; what requires an
explanation is how the motion can be both directional and highly correlated.
This physical information is introduced through one constraint that acts
simultaneously on all particles. The constraint involves a function $\phi
(x)=\phi (x_{1}\ldots x_{N})$ on configuration space $\mathbf{X}_{N}$ that
we call the \textquotedblleft drift\textquotedblright\ potential. We impose
that the displacements $\Delta x_{n}^{a}$ are such that the expected change
of the drift potential $\left\langle \Delta \phi \right\rangle $ is
constrained to be \ 
\begin{equation}
\left\langle \Delta \phi \right\rangle =\sum\limits_{n=1}^{N}\langle \Delta
x_{n}^{a}\rangle \frac{\partial \phi }{\partial x_{n}^{a}}=\kappa ^{\prime
}(x)~,  \label{constraint phi}
\end{equation}%
where $\kappa ^{\prime }(x)$ is another small but for now unspecified
function. As we shall later see this information is already sufficient to
construct an interesting ED. However, to reproduce the particular dynamics
that describes quantum systems we must further require that the potential $%
\phi (x)$ be a multi-valued function with the topological properties of an
angle --- $\phi $ and $\phi +2\pi $ represent the same angle.\footnote{%
The angular nature of the drift potential is explained when the ED framework
is extended to particles with spin \cite{Caticha Carrara 2019}.}

The physical origin of the drift potential $\phi (x)$ is at this point
unknown so how can one justify its introduction? The idea is that
identifying the relevant constraints can represent significant progress even
when their physical origin remains unexplained. Indeed, with the \emph{single%
}\ assumption of a constraint involving a drift potential we will explain
and coordinate \emph{several} features of quantum mechanics such as
entanglement, the existence of complex and symplectic structures, the actual
form of the Hamiltonian, and the linearity of the Schr\"{o}dinger equation.

\paragraph*{The gauge constraints ---}

The single constraint (\ref{constraint phi}) already leads to a rich
entropic dynamics but by imposing additional constraints we can construct
even more realistic models. To incorporate the effect of an external
electromagnetic field we impose that for each particle $n$ the expected
displacement $\langle \Delta x_{n}^{a}\rangle $ will satisfy%
\begin{equation}
\langle \Delta x_{n}^{a}\rangle A_{a}(x_{n})=\kappa _{n}^{\prime \prime
}\quad \text{for\quad }n=1\ldots N~,~  \label{constraint A}
\end{equation}%
where the electromagnetic vector potential $A_{a}(x_{n})$ is a field that
lives in the 3-dimensional physical space ($x_{n}\in \mathbf{X}$). The
strength of the coupling is given by the values of the $\kappa _{n}^{\prime
\prime }$. These quantities could be specified directly but, as is often the
case in entropic inference, it is much more convenient to specify them
indirectly in terms of the corresponding Lagrange multipliers.

\paragraph*{The transition probability ---}

An important feature of the ED model can already be discerned. The central
object of the discussion so far, the transition probability $P(x^{\prime
}|x) $, codifies information supplied through the prior and the constraints
which makes no reference to anything earlier than the initial position $x$.
Therefore ED must take the form of a Markov process.

The distribution $P(x^{\prime }|x)$ that maximizes the entropy $\mathcal{S}%
[P,Q]$ in (\ref{entropy a}) relative to (\ref{prior}) and subject to (\ref%
{constraint phi}), (\ref{constraint A}), and normalization is

\begin{equation}
P(x^{\prime }|x)=\frac{1}{Z}\exp -\tsum\nolimits_{n}\left( \frac{\alpha _{n}%
}{2}\delta _{ab}\Delta x_{n}^{a}\Delta x_{n}^{b}-\alpha ^{\prime }\left[
\partial _{na}\phi -\beta _{n}A_{a}(x_{n})\right] \Delta x_{n}^{a}\right) ~
\label{trans prob a}
\end{equation}%
where $\alpha ^{\prime }$ and $\beta _{n}$ are Lagrange multipliers. This is
conveniently written as

\begin{equation}
P(x^{\prime }|x)=\frac{1}{Z}\exp -\tsum\nolimits_{n}\frac{\alpha _{n}}{2}%
\delta _{ab}\left( \Delta x_{n}^{a}-\Delta \bar{x}_{n}^{a}\right) \left(
\Delta x_{n}^{b}-\Delta \bar{x}_{n}^{b}\right) ~,  \label{trans prob b}
\end{equation}%
with a suitably modified normalization and 
\begin{equation}
\Delta \bar{x}_{n}^{a}=\frac{\alpha ^{\prime }}{\alpha _{n}}\left[ \partial
_{na}\phi -\beta _{n}A_{a}(x_{n})\right] =\langle \Delta x_{n}^{a}\rangle ~.
\label{drift a}
\end{equation}%
A generic displacement is expressed as a drift plus a fluctuation, 
\begin{equation}
\Delta x_{n}^{a}=\langle \Delta x_{n}^{a}\rangle +\Delta w_{n}^{a}~,
\label{delta x a}
\end{equation}%
where%
\begin{equation}
\langle \Delta w_{n}^{a}\rangle =0\,,\quad \text{and}\quad \langle \Delta
w_{n}^{a}\Delta w_{n^{\prime }}^{b}\rangle =\frac{1}{\alpha _{n}}\delta
_{nn^{\prime }}\delta ^{ab}~,  \label{fluct a}
\end{equation}

The fact that the constraints (\ref{constraint phi}) and (\ref{constraint A}%
) are not independent --- both involve the same displacements $\langle
\Delta x_{n}^{a}\rangle $ --- has turned out to be significant. We can
already see in (\ref{trans prob a}) and (\ref{drift a}) that it leads to a
gauge symmetry. As we shall later see the vector potential $A_{a}$ will be
interpreted as the corresponding gauge connection field and the multipliers $%
\beta _{n}$ will be related to the electric charges through $\beta
_{n}=q_{n}/\hbar c$.

\section{Entropic time}

The task of iterating the short steps described by the transition
probability (\ref{trans prob b}) to predict motion over finite distances
leads us to introduce a book-keeping parameter $t$, to be called time, in
order to keep track of the accumulation of short steps. The construction of
time involves three ingredients: \textbf{(a)} we must specify what we mean
by an `instant'; \textbf{(b)} these instants must be ordered; and finally, 
\textbf{(c)} one must specify the interval $\Delta t$ between successive
instants --- one must define `duration'.\ 

Since the foundation for any theory of time is the theory of change, that
is, the dynamics, the notion of time constructed below will reflect the
inferential nature of entropic dynamics. Such a construction we will call
\textquotedblleft entropic\textquotedblright\ time \cite{Caticha 2010}.
Later we will return to the question of whether and how this
\textquotedblleft entropic\textquotedblright\ time is related to the
\textquotedblleft physical\textquotedblright\ time that is measured by
clocks.

\subsection{Time as an ordered sequence of instants}

ED consists of a succession of short steps. Consider, for example, the $i$th
step which takes the system from $x=x_{i-1}$ to $x^{\prime }=x_{i}$.
Integrating the joint probability, $P(x_{i},x_{i-1})$, over $x_{i-1}$ gives 
\begin{equation}
P(x_{i})=\int dx_{i-1}P(x_{i},x_{i-1})=\int
dx_{i-1}P(x_{i}|x_{i-1})P(x_{i-1})~.
\end{equation}%
No physical assumptions were involved in deriving this equation; it follows
directly from the laws of probability. To establish the connection to time
and dynamics we will make the physical assumption that if $P(x_{i-1})$ is
interpreted as the probability of different values of $x_{i-1}$ at one
\textquotedblleft instant\textquotedblright\ labelled\emph{\ }$t$, then we
will interpret $P(x_{i})$ as the probability of values of $x_{i}$ at the
next \textquotedblleft instant\textquotedblright\ labelled $t^{\prime }$.
More explicitly, if we write $P(x_{i-1})=\rho _{t}(x)$ and $P(x_{i})=\rho
_{t^{\prime }}(x^{\prime })$ then we have 
\begin{equation}
\rho _{t^{\prime }}(x^{\prime })=\int dx\,P(x^{\prime }|x)\rho _{t}(x)~.
\label{ED a}
\end{equation}%
This equation defines the notion of \textquotedblleft
instant\textquotedblright :\ if the distribution $\rho _{t}(x)$\ refers to
one instant $t$, then the distribution $\rho _{t^{\prime }}(x^{\prime })$\
generated by $P(x^{\prime }|x)$ defines what we mean by the
\textquotedblleft next\textquotedblright\ instant $t^{\prime }$. Iterating
this process defines the dynamics.

This construction of time is intimately related to information and
inference. An instant is an informational state that is complete in the
sense that it is specified by the information --- codified into the
distributions $\rho _{t}(x)$ and $P(x^{\prime }|x)$ --- that is sufficient
for predicting the next instant. Thus, \emph{the present is defined through
a sufficient amount of information such that,\ given the present, the future
is independent of the past}.

In the ED framework the notions of instant and of simultaneity are
intimately related to the distribution $\rho _{t}(x)$. To see how this comes
about consider a single particle at the point $\vec{x}=(x^{1},x^{2},x^{3})$.
It is implicit in the notation that $x^{1}$, $x^{2}$, and $x^{3}$ occur
simultaneously. When we describe a system of $N$ particles by a single point 
$x=(\vec{x}_{1},\vec{x}_{2},\ldots \vec{x}_{N})$ in $3N$-dimensional
configuration space it is also implicitly assumed that all the $3N$
coordinate values refer to the same instant; they are simultaneous. The very
idea of a \emph{point} in configuration space assumes simultaneity. And
furthermore, whether we deal with one particle or many, a distribution such
as $\rho _{t}(x)$ is meant to describe our uncertainty about the possible
configurations $x$ of the system at the given instant. Thus, a probability
distribution $\rho _{t}(x)$ provides a criterion of simultaneity.\footnote{%
In a relativistic theory there is more freedom in the choice of instants and
this translates into a greater flexibility with the notion of simultaneity.
Conversely, the requirement of consistency among the different notions of
simultaneity severely limits the allowed forms of relativistic ED \cite{Ipek
Abedi Caticha 2018}.}

\subsection{The arrow of entropic time}

The notion of time constructed according to eq.(\ref{ED a}) is intrinsically
directional. There is an absolute sense in which $\rho _{t}(x)$\ is prior
and $\rho _{t^{\prime }}(x^{\prime })$\ is posterior. Indeed, the same rules
of probability that led us to (\ref{ED a}) can also lead us to the
time-reversed evolution, 
\begin{equation}
\rho _{t}(x)=\int dx^{\prime }\,P(x|x^{\prime })\rho _{t^{\prime
}}(x^{\prime })\,.
\end{equation}%
Note, however, that there is a temporal asymmetry: while the distribution $%
P(x^{\prime }|x)$, eq.(\ref{trans prob a}), is a Gaussian derived using the
maximum entropy method, its time-reversed version $P(x|x^{\prime })$ is
related to $P(x^{\prime }|x)$ by Bayes' theorem, 
\begin{equation}
P(x|x^{\prime })=\frac{\rho _{t}(x)}{\rho _{t^{\prime }}(x^{\prime })}%
P(x^{\prime }|x)~,  \label{bt1}
\end{equation}%
which in general will not be Gaussian.

The puzzle of the arrow of time (see \emph{e.g.} \cite{Price 1996}\cite{Zeh
2007}) arises from the difficulty in deriving a temporal asymmetry from
underlying laws of nature that are symmetric. The ED approach offers a fresh
perspective on this topic because it does not assume any underlying laws of
nature --- whether they be symmetric or not. The asymmetry is the inevitable
consequence of constructing time in a dynamics driven by entropic inference.

From the ED point of view the challenge does not consist in explaining the
arrow of time --- \emph{entropic time itself only flows forward} --- but
rather in explaining how it comes about that despite the arrow of time some
laws of physics, such as the Schr\"{o}dinger equation, turn out to be time
reversible. We will revisit this topic in section \ref{entropic v physical
time}.

\subsection{Duration and the sub-quantum motion}

We have argued that the concept of time is intimately connected to the
associated dynamics but at this point neither the transition probability $%
P(x^{\prime }|x)$ that specifies the dynamics nor the corresponding entropic
time have been fully defined yet. It remains to specify how the multipliers $%
\alpha _{n}$ and $\alpha ^{\prime }$ are related to the interval $\Delta t$
between successive instants.

The basic criterion for this choice is convenience: \emph{duration is
defined so that motion looks simple. }The description of motion is simplest
when it reflects the symmetry of translations in space and time. In a flat
space-time this leads to an entropic time that resembles Newtonian time in
that it flows \textquotedblleft equably\ everywhere and
everywhen.\textquotedblright\ Referring to eqs.(\ref{drift a}) and (\ref%
{fluct a}) we choose $\alpha ^{\prime }$ and $\alpha _{n}$ to be independent
of $x$ and $t$, and we choose the ratio $\alpha ^{\prime }/\alpha
_{n}\propto \Delta t$ so that there is a well defined drift velocity. For
future convenience the proportionality constants will be expressed in terms
of some particle-specific constants $m_{n}$, 
\begin{equation}
\frac{\alpha ^{\prime }}{\alpha _{n}}=\frac{\hbar }{m_{n}}\Delta t~,
\label{alpha n}
\end{equation}%
where $\hbar $ is an overall constant that fixes the units of the $m_{n}$s
relative to the units of time. As we shall later see, the constants $m_{n}$
will eventually be identified with the particle masses while the constant $%
\hbar $ will be identified as Planck's constant. Having specified the ratio $%
\alpha ^{\prime }/\alpha _{n}$ it remains to specify $\alpha _{n}$ (or $%
\alpha ^{\prime }$). It turns out that the choice is not unique. There is a
variety of motions at the sub-quantum \textquotedblleft
microscopic\textquotedblright\ level that lead to the same quantum mechanics
at the \textquotedblleft macroscopic\textquotedblright\ level.

In previous work \cite{Caticha 2010}-\cite{Caticha 2017} we chose $\alpha
_{n}$\ proportional to $1/\Delta t$. This led to an ED in which the
particles follow the highly irregular non-differentiable Brownian
trajectories characteristic of an Einstein-Smoluchowski process. The first
new contribution of this paper is to explore the consequences of choosing $%
\alpha _{n}\propto 1/\Delta t^{3}$, 
\begin{equation}
\alpha _{n}=\frac{m_{n}}{\eta \Delta t^{3}}~,  \label{alpha prime}
\end{equation}%
where a new constant $\eta $ is introduced.

It is convenient to introduce a notation tailored to configuration space.
Let $x^{A}=x_{n}^{a}$, $\partial _{A}=\partial /\partial x_{n}^{a}$, and $%
\delta _{AB}=\delta _{nn^{\prime }}\delta _{ab}$, where the upper case
indices $A,B,\ldots $ label both the particles $n,n^{\prime },\ldots $ and
their coordinates $a,b,\ldots $ Then the transition probability (\ref{trans
prob b}) becomes 
\begin{equation}
P(x^{\prime }|x)=\frac{1}{Z}\exp \left[ -\frac{1}{2\eta \Delta t}%
m_{AB}\left( \frac{\Delta x^{A}}{\Delta t}-v^{A}\right) \left( \frac{\Delta
x^{B}}{\Delta t}-v^{B}\right) \right] ~,  \label{trans prob c}
\end{equation}%
where we used (\ref{drift a}) to define the drift velocity, 
\begin{equation}
v^{A}=\frac{\langle \Delta x^{A}\rangle }{\Delta t}=m^{AB}\left[ \partial
_{B}\Phi -\bar{A}_{B}\right] ~.  \label{drift velocity}
\end{equation}%
The drift potential is rescaled into a new variable 
\begin{equation}
\Phi =\hbar \phi  \label{phase a}
\end{equation}%
which will be called \emph{the phase}. We also introduced the
\textquotedblleft mass\textquotedblright\ tensor and its inverse, 
\begin{equation}
m_{AB}=m_{n}\delta _{AB}=m_{n}\delta _{nn^{\prime }}\delta _{ab}\quad \text{%
and}\quad m^{AB}=\frac{1}{m_{n}}\delta ^{AB}~,  \label{mass a}
\end{equation}%
and $\bar{A}_{A}$ is a field in configuration space with components, 
\begin{equation}
\bar{A}_{A}(x)=\hbar \beta _{n}A_{a}(x_{n})~,  \label{config space A}
\end{equation}%
A generic displacement is then written as a drift plus a fluctuation, 
\begin{equation}
\Delta x^{A}=v^{A}\Delta t+\Delta w^{A}~.  \label{delta x b}
\end{equation}%
and the fluctuations $\Delta w^{A}$ are given by 
\begin{equation}
\langle \Delta w^{A}\rangle =0\quad \text{and}\quad \langle \Delta
w^{A}\Delta w^{B}\rangle =\eta m^{AB}\Delta t^{3}~,  \label{fluct b}
\end{equation}%
or 
\begin{equation}
\left\langle \left( \frac{\Delta x^{A}}{\Delta t}-v^{A}\right) \left( \frac{%
\Delta x^{B}}{\Delta t}-v^{B}\right) \right\rangle =\eta m^{AB}\Delta t.
\label{fluct c}
\end{equation}%
It is noteworthy that $\langle \Delta x^{A}\rangle \sim O(\Delta t)$ and $%
\Delta w^{A}\sim O(\Delta t^{3/2})$. This means that for short steps the
fluctuations are negligible and the dynamics is dominated by the drift. The
particles follow trajectories that are indeterministic but differentiable.
Since $\Delta w^{A}\sim O(\Delta t^{3/2})$ the limit 
\begin{equation}
V^{A}=\lim_{\Delta t\rightarrow 0}\frac{\Delta x^{A}}{\Delta t}=v^{A}~.
\label{velocity a}
\end{equation}%
is well defined. In words: the actual velocities of the particles coincide
with the expected or drift velocities. From eq.(\ref{drift velocity}) we see
that these velocities are continuous functions. The question of whether the
velocities themselves are differentiable or not is trickier.

Consider two successive displacements $\Delta x=x^{\prime }-x$ followed by $%
\Delta x^{\prime }=x^{\prime \prime }-x^{\prime }$. The velocities are 
\begin{equation}
V^{A}=\frac{\Delta x^{A}}{\Delta t}\quad \text{and}\quad V^{\prime A}=\frac{%
\Delta x^{\prime A}}{\Delta t}
\end{equation}%
The change in velocity is given by a Langevin equation, 
\begin{equation}
\Delta V^{A}=\langle \Delta V^{A}\rangle _{x^{\prime \prime }x^{\prime
}}+\Delta U^{A}
\end{equation}%
where $\langle \cdot \rangle _{x^{\prime \prime }x^{\prime }}$ denotes
taking the expectations over $x^{\prime \prime }$ using $P(x^{\prime \prime
}|x^{\prime })$, and then over $x^{\prime }$ using $P(x^{\prime }|x)$, and $%
\Delta U^{A}$ is a fluctuation. It is straightforward to show that 
\begin{equation}
\langle \Delta V^{A}\rangle _{x^{\prime \prime }x^{\prime }}=(\partial
_{t}v^{A}+v^{B}\partial _{B}v^{A})\,\Delta t~,
\end{equation}%
so that the expected acceleration is given by the convective derivative of
the velocity field along itself, 
\begin{equation}
\lim_{\Delta t\rightarrow 0}\frac{\langle \Delta V^{A}\rangle }{\Delta t}%
=(\partial _{t}+v^{B}\partial _{B})v^{A}~.
\end{equation}%
One can also show that

\begin{equation}
\langle \Delta U^{A}\rangle _{x^{\prime \prime }x^{\prime }}=0~,\quad \text{%
and}\quad \langle \Delta U^{A}\Delta U^{B}\rangle _{x^{\prime \prime
}x^{\prime }}=2\eta m^{AB}\Delta t~,  \label{fluct d}
\end{equation}%
which means that $\Delta U$ is a Wiener process and we deal with a Brownian
motion of the Oernstein-Uhlenbeck type.

We conclude this section with some general remarks.

\subparagraph*{On the nature of clocks ---}

In Newtonian mechanics time is defined to simplify the dynamics. The
prototype of a clock is a free particle which moves equal distances in equal
times. In ED time is also defined to simplify the dynamics of free particles
(for sufficiently short times all particles are free) and the prototype of a
clock is a free particle too: as we see in (\ref{delta x b}) \emph{the
particle's mean displacement increases by equal amounts in equal times}.

\subparagraph*{On the nature of mass ---}

In standard quantum mechanics, \textquotedblleft what is
mass?\textquotedblright\ and \textquotedblleft why quantum
fluctuations?\textquotedblright\ are two independent mysteries. In ED the
mystery is somewhat alleviated: as we see in eq.(\ref{fluct c}) mass and
fluctuations are two sides of the same coin. \emph{Mass is an inverse
measure of the velocity fluctuations}.

\subparagraph*{The information metric of configuration space ---}

In addition to defining the dynamics the transition probability eq.(\ref%
{trans prob c}) serves to define the geometry of the $N$-particle
configuration space, $\mathbf{X}_{N}$. Since the physical single particle
space $\mathbf{X}$ is described by the Euclidean metric $\delta _{ab}$ we
can expect that the $N$-particle configuration space, $\mathbf{X}_{N}=%
\mathbf{X}\times \ldots \times \mathbf{X}$, will also be flat, but for
non-identical particles a question might be raised about the relative scales
or weights associated to each $\mathbf{X}$ factor. Information geometry
provides the answer.

The fact that to each point $x\in \mathbf{X}_{N}$ there corresponds a
probability distribution $P(x^{\prime }|x)$ means that to the space $\mathbf{%
X}_{N}$ we can associate a statistical manifold the geometry of which (up to
an overall scale factor) is uniquely determined by the information metric 
\cite{Caticha 2012}\cite{Amari 1985}, 
\begin{equation}
\gamma _{AB}=\int dx^{\prime }\,P(x^{\prime }|x)\frac{\partial \log
P(x^{\prime }|x)}{\partial x^{A}}\frac{\partial \log P(x^{\prime }|x)}{%
\partial x^{B}}~.  \label{gamma C}
\end{equation}%
Substituting eqs.(\ref{trans prob c}) into (\ref{gamma C}) yields 
\begin{equation}
\gamma _{AB}=\frac{1}{\eta \Delta t^{3}}m_{AB}~.  \label{gamma AB}
\end{equation}%
The divergence as $\Delta t\rightarrow 0$ arises because the information
metric measures statistical distinguishability. As $\Delta t\rightarrow 0$
the distributions $P(x^{\prime }|x)$ and $P(x^{\prime }|x+\Delta x)$ become
more sharply peaked and increasingly easier to distinguish so that $\gamma
_{AB}\rightarrow \infty $. Thus, up to a scale factor the metric of
configuration space is basically the mass tensor.

The practice of describing a many-particle system as a single point in an
abstract configuration space goes back to the work of H. Hertz in 1894 \cite%
{Lanczos 1970}. Historically the choice of the mass tensor as the metric of
configuration space has been regarded as being convenient but of no
particular significance. We can now see that the choice is not just a merely
useful convention: up to an overall scale the metric follows uniquely from
information geometry. Furthermore, it suggests the intriguing possibility of
a deeper connection between kinetic energy and information geometry.

\subparagraph*{Invariance under gauge transformations ---}

The fact that constraints (\ref{constraint phi}) and (\ref{constraint A})
are not independent --- they are both linear in the same displacements $%
\langle \Delta x_{n}^{a}\rangle $ --- leads to a gauge symmetry. This is
evident in eq.(\ref{trans prob a}) where $\phi $ and $A_{a}$ appear in the
combination $\partial _{na}\phi -\beta _{n}A_{a}$ which is invariant under
the gauge transformations, 
\begin{eqnarray}
A_{a}(x_{n}) &\rightarrow &A_{a}^{\prime }(x_{n})=A_{a}(x_{n})+\partial
_{a}\chi (x_{n})~, \\
\phi (x) &\rightarrow &\phi ^{\prime }(x)=\phi (x)+\dsum\nolimits_{n}\beta
_{n}\chi (x_{n})~.
\end{eqnarray}%
These transformations are local in 3d-space. Introducing 
\begin{equation}
\bar{\chi}(x)=\dsum\nolimits_{n}\hbar \beta _{n}\chi (x_{n})~,
\label{config space chi}
\end{equation}%
they can be written in the $N$-particle configuration space, 
\begin{eqnarray}
\bar{A}_{A}(x) &\rightarrow &\bar{A}_{A}^{\prime }(x)=\bar{A}%
_{A}(x)+\partial _{A}\bar{\chi}(x)~, \\
\Phi (x) &\rightarrow &\Phi ^{\prime }(x)=\Phi (x)+\bar{\chi}(x)~.
\label{GT phase}
\end{eqnarray}

\noindent \textbf{Interpretation:} The drift potential $\phi (x)=\phi (\vec{x%
}_{1},\vec{x}_{2},\ldots )$ is assumed to be an \textquotedblleft
angle\textquotedblright\ --- $\phi (x)$ and $\phi (x)+2\pi $ are meant to
describe the same angle. The angle at $\vec{x}_{1}$ depends on the values of
all the other positions $\vec{x}_{2},\vec{x}_{3},\ldots $, and the angle at $%
\vec{x}_{2}$ depends on the values of all the other positions $\vec{x}_{1},%
\vec{x}_{3},\ldots $, and so on. The fact that the origins from which these
angles are measured can be redefined by different amounts at different
places gives rise to a local gauge symmetry. In order to compare angles at
different locations one introduces a \emph{connection} field, the vector
potential $A_{a}(\vec{x})$. It defines which origin at $\vec{x}+\Delta \vec{x%
}$ is the \textquotedblleft same\textquotedblright\ as the origin at $\vec{x}
$. This is implemented by imposing that as we change origins and $\Phi (x)$
changes to $\Phi +\bar{\chi}$ then the connection transforms as $%
A_{a}\rightarrow A_{a}+\partial _{a}\chi $ so that the quantity $\partial
_{A}\Phi -\bar{A}_{A}$ remains invariant.

\subparagraph*{A fractional Brownian motion? ---}

The choices $\alpha _{n}\varpropto 1/\Delta t$ and $\alpha _{n}\varpropto
1/\Delta t^{3}$ lead to Einstein-Smoluchowski and Oernstein-Uhlenbeck
processes respectively. For definiteness throughout the rest of this paper
we will assume that the sub-quantum motion is an OU process but more general
fractional Brownian motions \cite{Mandelbrot Van Ness 1968} are possible.
Consider 
\begin{equation}
\alpha _{n}=\frac{m_{n}}{\eta \Delta t^{\gamma }}~,
\end{equation}%
where $\gamma \ $is some positive parameter. The corresponding transition
probability (\ref{trans prob b}), 
\begin{equation}
P(x^{\prime }|x)=\frac{1}{Z}\exp \left[ -\frac{1}{2\eta \Delta t^{\gamma }}%
m_{AB}\left( \Delta x^{A}-v^{A}\Delta t\right) \left( \Delta
x^{B}-v^{B}\Delta t\right) \right] \ ,
\end{equation}%
leads to fluctuations such that%
\begin{equation}
\langle \Delta w^{A}\rangle =0\quad \text{and}\quad \langle \Delta
w^{A}\Delta w^{B}\rangle =\eta m^{AB}\Delta t^{\gamma }~,
\end{equation}%
or 
\begin{equation}
\left\langle \left( \frac{\Delta x^{A}}{\Delta t}-v^{A}\right) \left( \frac{%
\Delta x^{B}}{\Delta t}-v^{B}\right) \right\rangle =\eta m^{AB}\Delta
t^{\gamma -2}.
\end{equation}%
We will not pursue this topic further except to note that since $\langle
\Delta x^{A}\rangle \sim O(\Delta t)$ and $\Delta w^{A}\sim O(\Delta
t^{\gamma /2})$ for $\gamma <2$ the sub-quantum motion is dominated by
fluctuations and the trajectories are non-differentiable, while for $\gamma
>2$ the drift dominates and velocities are well defined.

\section{The evolution equation in differential form}

Entropic dynamics is generated by iterating eq.(\ref{ED a}): given the
information that defines one instant, the integral eq.(\ref{ED a}) is used
to construct the next instant. As so often in physics it is more convenient
to rewrite the equation of evolution in differential form. The result is 
\begin{equation}
\partial _{t}\rho =-\partial _{A}\left( v^{A}\rho \right) ~,  \label{CE}
\end{equation}%
where $v^{A}$ is given by (\ref{drift velocity}). Before we proceed to its
derivation we note that eq.(\ref{CE}) is a consequence of the fact that the
particles follow continuous paths. Accordingly, we will follow standard
practice and call it \emph{the continuity equation}. Also note that in the
OU process considered here ($\gamma =3$) the current velocity --- the
velocity with which the probability flows in configuration space ---
coincides with the drift velocity (\ref{drift velocity}) and with the actual
velocities of the particles (\ref{velocity a}).\footnote{%
In the ES type of ED considered in previous papers ($\gamma =1$) \cite%
{Caticha 2010}-\cite{Caticha 2017} the probability also satisfies a
continuity equation --- a Fokker-Planck equation --- and the current
velocity is the sum of the drift velocity plus an osmotic component 
\begin{equation*}
u^{A}=-\hbar m^{AB}\partial _{B}\log \rho ^{1/2}
\end{equation*}%
due to diffusion.}

Next we derive (\ref{CE}) using a technique that is well known in diffusion
theory \cite{Chandrasekar 1943}. (For an alternative derivation see \cite%
{DiFranzo 2018}.) The result of building up a \emph{finite} change from an
initial time $t_{0}$ to a later time $t$ leads to the distribution%
\begin{equation}
\rho (x,t)=\int dx_{0}\,P(x,t|x_{0},t_{0})\rho (x_{0},t_{0})~,  \label{CK c}
\end{equation}%
where the finite-time transition probability, $P(x,t|x_{0},t_{0})$, is
constructed by iterating the infinitesimal changes described in eq.(\ref{ED
a}), 
\begin{equation}
P(x,t+\Delta t|x_{0},t_{0})=\int dz\,P(x,t+\Delta t|z,t)P(z,t|x_{0},t_{0})~.
\label{CK d}
\end{equation}%
For small times $\Delta t$ the distribution $P(x,t+\Delta t|z,t)$, given in
eq. (\ref{trans prob c}), is very sharply peaked at $x=z$. In fact, as $%
\Delta t\rightarrow 0$ we have $P(x,t+\Delta t|z,t)\rightarrow \delta (x-z)$%
. Such singular behavior cannot be handled directly by Taylor expanding in $%
z $ about the point $x$. Instead one follows an indirect procedure. Multiply
by a smooth test function $f(x)$ and integrate over $x$, 
\begin{equation}
\int dx\,P(x,t+\Delta t|x_{0},t_{0})f(x)=\int dz\left[ \int dx\,P(x,t+\Delta
t|z,t)f(x)\right] P(z,t|x_{0},t_{0})~.{}  \label{CK e}
\end{equation}%
The test function $f(x)$ is assumed sufficiently smooth precisely so that it
can be expanded about $z$. Then as $\Delta t\rightarrow 0$ the integral in
the brackets, dropping all terms of order higher than $\Delta t$, is 
\begin{align}
\left[ \cdots \right] & =\int dx\,P(x,t+\Delta t|z,t)\left( f(z)+\frac{%
\partial f}{\partial z^{A}}(x^{A}-z^{A})+...\right)  \notag \\
& =f(z)+v^{A}(z)\Delta t\frac{\partial f}{\partial z^{A}}+\ldots
\label{CK f}
\end{align}%
where we used eq.(\ref{delta x b}). Next substitute (\ref{CK f}) into the
right hand side of (\ref{CK e}), divide by $\Delta t$, and let $\Delta
t\rightarrow 0$. Since $f(x)$ is arbitrary the result is 
\begin{equation}
\partial _{t}P(x,t|x_{0},t_{0})=-\partial
_{A}[v^{A}(x)P(x,t|x_{0},t_{0})]~\,,  \label{CE b}
\end{equation}%
which is the continuity equation for the finite-time transition probability.
Differentiating eq.(\ref{CK c}) with respect to $t$, and substituting (\ref%
{CE b}) completes the derivation of the continuity equation (\ref{CE}).

The continuity equation (\ref{CE}) can be written in another equivalent but
very suggestive form involving functional derivatives. For some suitably
chosen functional $\tilde{H}[\rho ,\Phi ]$ we have 
\begin{equation}
\partial _{t}\rho (x)=-\partial _{A}\left[ \rho m^{AB}(\partial _{B}\Phi -%
\bar{A}_{B})\right] =\frac{\delta \tilde{H}}{\delta \Phi (x)}~.
\label{Hamilton a}
\end{equation}%
It is easy to check that the appropriate functional $\tilde{H}$ is%
\begin{equation}
\tilde{H}[\rho ,\Phi ]=\int dx\,\frac{1}{2}\rho m^{AB}\left( \partial
_{A}\Phi -\bar{A}_{A}\right) \left( \partial _{B}\Phi -\bar{A}_{B}\right)
+F[\rho ]~,  \label{Hamiltonian a}
\end{equation}%
where the unspecified functional $F[\rho ]$ is an integration constant.%
\footnote{%
Eqs.(\ref{Hamilton a}) and (\ref{Hamiltonian a}) show the reason to have
introduced the new variable $\Phi =\hbar \phi $. With this choice $\Phi $
will eventually be recognized as the momentum that is canonically conjugate
to the generalized coordinate $\rho $ with Hamiltonian $\tilde{H}$.}

The continuity equation (\ref{Hamilton a}) describes a somewhat peculiar OU
Brownian motion in which the probability density $\rho (x)$ is driven by the
non-dynamical fields\ $\Phi $, and $\bar{A}$. This is an interesting ED in
its own right but it is not QM. Indeed, a \emph{quantum} dynamics consists
in the coupled evolution of two dynamical fields: the density $\rho (x)$ and
the phase of the wave function. This second field can be naturally
introduced into ED by allowing the phase field $\Phi $ in (\ref{drift
velocity}) to become dynamical which amounts to an ED in which the
constraint (\ref{constraint phi}) is continuously updated at each instant in
time. Our next topic is to propose the appropriate updating criterion. It
yields an ED in which the phase field $\Phi $ guides the evolution of $\rho $%
, and in return, the evolving $\rho $ reacts back and induces a change in $%
\Phi $.

\section{The epistemic phase space}

In ED we deal with two configuration spaces. One is the \emph{ontic
configuration space} $\mathbf{X}_{N}=\mathbf{X}\times \mathbf{X}\times
\ldots $ of all particle positions, $x=(x_{1}\ldots x_{N})\in \mathbf{X}_{N}$%
. The other is the \emph{epistemic configuration space }or\emph{\
e-configuration space} $\mathbf{P}$ of all normalized probabilities, 
\begin{equation}
\mathbf{P}=\left\{ \rho \left\vert \rho (x)\geq 0;\int dx\rho (x)=1\right.
\right\} ~.  \label{e-config space}
\end{equation}%
To formulate the coupled dynamics of $\rho $ and $\Phi $ we need a framework
to study paths in the larger space $\{\rho ,\Phi \}$ that we will call the 
\emph{epistemic phase space} or \emph{e-phase space}.

Given any manifold such as $\mathbf{P}$ the associated tangent and cotangent
bundles, respectively $T\mathbf{P}$ and $T^{\ast }\mathbf{P}$, are geometric
objects that are always available to us independently of any physical
considerations. Both are manifolds in their own right but the cotangent
bundle $T^{\ast }\mathbf{P}$ --- the space of all probabilities and all
covectors --- is of particular interest because it comes automatically
endowed with a rich geometrical structure \cite{Kibble 1979}-\cite{Ashtekar
Schilling 1998}.\footnote{%
We deal with $\infty $-dimensional spaces. The level of mathematical rigor
in what follows is typical of theoretical physics --- which is a euphemism
for \textquotedblleft from very low to none at all.\textquotedblright\ For a
more sophisticated treatment see \cite{Cirelli et al 1990}\cite{Ashtekar
Schilling 1998}.} The point is that cotangent bundles are symplectic
manifolds and this singles out as \textquotedblleft
natural\textquotedblright\ those dynamical laws that happen to preserve some
privileged symplectic form. This observation will lead us to identify
e-phase space $\{\rho ,\Phi \}$\ with the cotangent bundle\emph{\ }$T^{\ast }%
\mathbf{P}$ and provides the natural criterion for updating constraints,
that is, for updating the phase $\Phi $.

\subsection{Notation: vectors, covectors, etc.}

A point $X\in T^{\ast }\mathbf{P}$ will be represented as 
\begin{equation}
X=\left( \rho (x),\pi (x)\right) =(\rho ^{x},\pi _{x})~,
\end{equation}%
where $\rho ^{x}$ represents coordinates on the base manifold $\mathbf{P}$
and $\pi _{x}$ represents some generic coordinates on the space $T^{\ast }%
\mathbf{P}_{\rho }$ that is cotangent to $\mathbf{P}$ at the point $\rho $.
Curves in $T^{\ast }\mathbf{P}$ allow us to define vectors. Let $X=X(\lambda
)$ be a curve parametrized by $\lambda $, then the vector $\bar{V}$ tangent
to the curve at $X=(\rho ,\pi )$ has components $d\rho ^{x}/d\lambda $ and $%
d\pi _{x}/d\lambda $, and is written 
\begin{equation}
\bar{V}=\frac{d}{d\lambda }=\int dx\,\left[ \frac{d\rho ^{x}}{d\lambda }%
\frac{\delta }{\delta \rho ^{x}}+\frac{d\pi _{x}}{d\lambda }\frac{\delta }{%
\delta \pi _{x}}\right] ~,  \label{vector}
\end{equation}%
where $\delta /\delta \rho ^{x}$ and $\delta /\delta \pi _{x}$ are the basis
vectors. The directional derivative of a functional $F[X]$ along the curve $%
X(\lambda )$ is 
\begin{equation}
\frac{dF}{d\lambda }=\mathsf{\tilde{\nabla}}F[\bar{V}]=\int dx\,\left[ \frac{%
\delta F}{\delta \rho ^{x}}\frac{d\rho ^{x}}{d\lambda }+\frac{\delta F}{%
\delta \pi _{x}}\frac{d\pi _{x}}{d\lambda }\right] ~,  \label{dir deriv}
\end{equation}%
where $\mathsf{\tilde{\nabla}}$ is the functional gradient in $T^{\ast }%
\mathbf{P}$, that is, the gradient of a generic functional $F[X]=F[\rho ,\pi
]$ is 
\begin{equation}
\mathsf{\tilde{\nabla}}F=\int dx\,\left[ \frac{\delta F}{\delta \rho ^{x}}%
\mathsf{\tilde{\nabla}}\rho ^{x}+\frac{\delta F}{\delta \pi _{x}}\mathsf{%
\tilde{\nabla}}\pi _{x}\right] ~.  \label{gradient}
\end{equation}%
The tilde `$\symbol{126}$' serves to distinguish the functional gradient $%
\tilde{\nabla}$ from the spatial gradient $\nabla f=\partial _{a}f\nabla
x^{a}$ on $\mathbf{R}^{3}$.

The fact that the space $\mathbf{P}$ is constrained to normalized
probabilities means that the coordinates $\rho ^{x}$ are not independent.
This technical difficulty is handled by embedding the $\infty $-dimensional
manifold $\mathbf{P}$ in a $(\infty +1)$-dimensional manifold $\mathbf{P}%
^{+1}$ where the coordinates $\rho ^{x}$ are unconstrained.\footnote{%
At this point the act of embedding $\mathbf{P}$ into $\mathbf{P}^{+1}$
represents no loss of generality because the embedding space $\mathbf{P}%
^{+1} $ remains unspecified.} Thus, strictly, $\mathsf{\tilde{\nabla}}F$ is
a covector on $T^{\ast }\mathbf{P}^{+1}$, that is, $\mathsf{\tilde{\nabla}}%
F\in T^{\ast }\left( T^{\ast }\mathbf{P}^{+1}\right) _{X}$ and $\mathsf{%
\tilde{\nabla}}\rho ^{x}$ and $\mathsf{\tilde{\nabla}}\pi _{x}$ are the
corresponding basis covectors. Nevertheless, the gradient $\mathsf{\tilde{%
\nabla}}F$ will yield the desired directional derivatives (\ref{dir deriv})
on $T^{\ast }\mathbf{P}$ provided its action is restricted to vectors $\bar{V%
}$ that are tangent to the manifold $\mathbf{P}$. Such tangent vectors are
constrained to obey 
\begin{equation}
\frac{d}{d\lambda }\int dx\rho (x)\,=\int dx\,\frac{d\rho ^{x}}{d\lambda }%
=0~.  \label{tang vector}
\end{equation}

Instead of keeping separate track of the $\rho ^{x}$ and $\pi _{x}$
coordinates it is more convenient to combine them into a single index. A
point $X=(\rho ,\pi )$ will then be labelled by its coordinates 
\begin{equation}
X^{I}=(X^{1x},X^{2x})=\left( \rho ^{x},\pi _{x}\right) ~.
\end{equation}%
We will use capital letters from the middle of the Latin alphabet ($%
I,J,K\ldots $); $I=(\alpha ,x)$ is a composite index where $\alpha =1,2$
keeps track of whether $x$ is an upper index ($\alpha =1$) or a lower index (%
$\alpha =2$).\footnote{%
This allows us, among other things, the freedom to switch from $\rho ^{x}$
to $\rho _{x}$ as convenience dictates; from now on $\rho _{x}=\rho
^{x}=\rho (x)$.} Then eqs.(\ref{vector}-\ref{gradient}) are written as%
\begin{equation}
\bar{V}=V^{I}\frac{\delta }{\delta X^{I}}~,\quad \text{where}\quad V^{I}=%
\frac{dX^{I}}{d\lambda }=%
\begin{bmatrix}
d\rho ^{x}/d\lambda \\ 
d\pi _{x}/d\lambda%
\end{bmatrix}%
~,
\end{equation}%
\begin{equation}
\frac{dF}{d\lambda }=\mathsf{\tilde{\nabla}}F[\bar{V}]=\frac{\delta F}{%
\delta X^{I}}V^{I}\quad \text{and}\quad \mathsf{\tilde{\nabla}}F=\,\frac{%
\delta F}{\delta X^{I}}\mathsf{\tilde{\nabla}}X^{I}~,
\end{equation}%
where the repeated indices indicate a summation over $\alpha $ and an
integration over $x$.

\subsection{The symplectic form in ED}

In classical mechanics with configuration space $\{q^{i}\}$ the Lagrangian $%
L(q,\dot{q})$ is a function on the tangent bundle while the Hamiltonian $%
H(q,p)$ is a function on the cotangent bundle \cite{Arnold 1997}\cite{Schutz
1980}. A symplectic form provides a mapping from the tangent to the
cotangent bundles. Given a Lagrangian the map is defined by $p_{i}=\partial
L/\partial \dot{q}^{i}$ and this automatically defines the corresponding
symplectic form. In ED there is no Lagrangian so in order to define the
symplectic map we must look elsewhere. We propose that the role played by
the Lagrangian in classical mechanics will in ED be played by the continuity
equation (\ref{Hamilton a}).

The fact that the preservation of a symplectic structure must reproduce the
continuity equation leads us to identify the phase $\Phi _{x}$ as the
momentum canonically conjugate to $\rho ^{x}$. This identification of the
e-phase space $\{\rho ,\Phi \}$ with $T^{\ast }\mathbf{P}$ is highly
non-trivial. It amounts to asserting that the phase $\Phi _{x}$ transforms
as the components of a Poincare 1-form%
\begin{equation}
\theta =\int dx\,\Phi _{x}\mathsf{d}\rho ^{x}~,
\end{equation}%
(where $\mathsf{d}$ is the exterior derivative) and the corresponding
symplectic 2-form $\Omega =-\mathsf{d}\theta $ is 
\begin{equation}
\Omega =\int dx\,\mathsf{d}\rho ^{x}\wedge \mathsf{d}\Phi _{x}=\int dx\,%
\left[ \tilde{\nabla}\rho ^{x}\otimes \tilde{\nabla}\Phi _{x}-\tilde{\nabla}%
\Phi _{x}\otimes \tilde{\nabla}\rho ^{x}\right] ~.  \label{sympl form a}
\end{equation}%
By construction $\Omega $ is exact ($\Omega =-\mathsf{d}\theta $) and closed
($\mathsf{d}\Omega =0$). The action of $\Omega \lbrack \cdot ,\cdot ]$ on
two vectors $\bar{V}=d/d\lambda $ and $\bar{U}=d/d\mu $ is given by 
\begin{equation}
\Omega \lbrack \bar{V},\bar{U}]=\int dx\,\left[ V^{1x}U^{2x}-V^{2x}U^{1x}%
\right] =\Omega _{IJ}V^{I}U^{J}~,~  \label{sympl form b}
\end{equation}%
so that the components of $\Omega $ are 
\begin{equation}
\Omega _{IJ}=\Omega _{\alpha x,\beta x^{\prime }}=%
\begin{bmatrix}
0 & 1 \\ 
-1 & 0%
\end{bmatrix}%
\delta (x,x^{\prime })~.  \label{sympl form c}
\end{equation}

\subsection{Hamiltonian flows and Poisson brackets}

Next we reproduce the $\infty $-dimensional $T^{\ast }\mathbf{P}$ analogues
of results that are standard in finite-dimensional classical mechanics \cite%
{Arnold 1997}\cite{Schutz 1980}. Given a vector field $\bar{V}[X]$ in
e-phase space we can integrate $V^{I}[X]=dX^{I}/d\lambda $ to find its
integral curves $X^{I}=X^{I}(\lambda )$. We are particularly interested in
those vector fields that generate flows that preserve the symplectic
structure, 
\begin{equation}
\pounds _{V}\Omega =0~,
\end{equation}%
where the Lie derivative is given by 
\begin{equation}
(\pounds _{V}\Omega )_{IJ}=V^{K}\tilde{\nabla}_{K}\Omega _{IJ}+\Omega _{KJ}%
\tilde{\nabla}_{I}V^{K}+\Omega _{IK}\tilde{\nabla}_{J}V^{K}~.
\end{equation}%
Since by eq.(\ref{sympl form c}) the components $\Omega _{IJ}$ are constant, 
$\tilde{\nabla}_{K}\Omega _{IJ}=0$, we can rewrite $\pounds _{V}\Omega $ as 
\begin{equation}
(\pounds _{V}\Omega )_{IJ}=\tilde{\nabla}_{I}(\Omega _{KJ}V^{K})-\tilde{%
\nabla}_{J}(\Omega _{KI}V^{K})~,
\end{equation}%
which is the exterior derivative (basically, the curl) of the covector $%
\Omega _{KI}V^{K}$. By Poincare's lemma, requiring $\pounds _{V}\Omega =0$
(a vanishing curl) implies that $\Omega _{KI}V^{K}$ is the gradient of a
scalar function, which we will denote $\tilde{V}[X]$,%
\begin{equation}
\Omega _{KI}V^{K}=\tilde{\nabla}_{I}\tilde{V}~.  \label{grad V}
\end{equation}%
Using (\ref{sympl form c}) this is more explicitly written as%
\begin{equation}
\int dx\,\left[ \frac{d\rho ^{x}}{d\lambda }\tilde{\nabla}\Phi _{x}-\frac{%
d\Phi _{x}}{d\lambda }\tilde{\nabla}\rho ^{x}\right] =\int dx\,\left[ \frac{%
\delta \tilde{V}}{\delta \rho ^{x}}\mathsf{\tilde{\nabla}}\rho ^{x}+\frac{%
\delta \tilde{V}}{\delta \Phi _{x}}\mathsf{\tilde{\nabla}}\Phi _{x}\right] ~,
\end{equation}%
or 
\begin{equation}
\frac{d\rho ^{x}}{d\lambda }=\frac{\delta \tilde{V}}{\delta \Phi _{x}}\quad 
\text{and}\quad \frac{d\Phi _{x}}{d\lambda }=-\frac{\delta \tilde{V}}{\delta
\rho ^{x}}~,  \label{Hamiltonian flow a}
\end{equation}%
which we recognize as Hamilton's equations for a Hamiltonian function $%
\tilde{V}$. This justifies calling $\bar{V}$ the Hamiltonian vector vector
field associated to the Hamiltonian function $\tilde{V}$.

From (\ref{sympl form b}), the action of the symplectic form $\Omega $ on
two Hamiltonian vector fields $\bar{V}=d/d\lambda $ and $\bar{U}=d/d\mu $
generated respectively by $\tilde{V}$ and $\tilde{U}$ is 
\begin{equation}
\Omega \lbrack \bar{V},\bar{U}]=\int dx\,\left[ \frac{d\rho ^{x}}{d\lambda }%
\frac{d\Phi _{x}}{d\mu }-\frac{d\Phi _{x}}{d\lambda }\frac{d\rho ^{x}}{d\mu }%
\right] ~,
\end{equation}%
which, using (\ref{Hamiltonian flow a}), gives%
\begin{equation}
\Omega \lbrack \bar{V},\bar{U}]=\int dx\,\left[ \frac{\delta \tilde{V}}{%
\delta \rho ^{x}}\frac{\delta \tilde{U}}{\delta \Phi _{x}}-\frac{\delta 
\tilde{V}}{\delta \Phi _{x}}\frac{\delta \tilde{U}}{\delta \rho ^{x}}\right] 
\overset{\text{def}}{=}\{\tilde{V},\tilde{U}\}~,  \label{PB a}
\end{equation}%
where on the right we introduced the Poisson bracket notation.

To summarize these results: (1) The condition for a flow generated by the
vector field $V^{I}$ to preserve the symplectic structure, $\pounds %
_{V}\Omega =0$, is that $V^{I}$ be the Hamiltonian vector field associated
to a Hamiltonian function $\tilde{V}$, eq.(\ref{Hamiltonian flow a}), 
\begin{equation}
V^{I}=\frac{dX^{I}}{d\lambda }=\{X^{I},\tilde{V}\}~.
\label{Hamiltonian flow b}
\end{equation}%
(2) The action of $\Omega $ on two Hamiltonian vector fields (\ref{PB a}) is
the Poisson bracket of the associated Hamiltonian functions,%
\begin{equation}
\Omega \lbrack \bar{V},\bar{U}]=\Omega _{IJ}V^{I}U^{J}=\{\tilde{V},\tilde{U}%
\}~.  \label{PB b}
\end{equation}

We conclude that the ED that preserves the symplectic structure $\Omega $
and reproduces the continuity equation (\ref{Hamilton a}) is described by
the Hamiltonian flow of the scalar functional $\tilde{H}$ in (\ref%
{Hamiltonian a}). However, the full dynamics, which will obey the
Hamiltonian evolution equations 
\begin{equation}
\partial _{t}\rho ^{x}=\frac{\delta \tilde{H}}{\delta \Phi _{x}}\quad \text{%
and}\quad \partial _{t}\Phi _{x}=-\frac{\delta \tilde{H}}{\delta \rho ^{x}}%
~,~  \label{Hamilton eqs a}
\end{equation}%
is not yet fully determined because the integration constant $F[\rho ]$ in (%
\ref{Hamiltonian a}) remains to be specified.

\subsection{The normalization constraint}

\label{normalization constraint}Since the particular flow that we will
associate with time evolution is required to reproduce the continuity
equation it will also preserve the normalization constraint, 
\begin{equation}
\tilde{N}=0\quad \text{where}\quad \tilde{N}=1-\left\vert \rho \right\vert
\quad \text{and}\quad \left\vert \rho \right\vert \overset{\text{def}}{=}%
\int dx\,\rho (x)~.  \label{N constraint}
\end{equation}%
Indeed, one can check that 
\begin{equation}
\partial _{t}\tilde{N}=\{\tilde{N},\tilde{H}\}=0~.  \label{N conservation}
\end{equation}

The Hamiltonian flow (\ref{Hamiltonian flow b}) generated by $\tilde{N}$ and
parametrized by $\alpha $ is given by the vector field 
\begin{equation}
\bar{N}=N^{I}\frac{\delta }{\delta X^{I}}\quad \text{with}\quad N^{I}=\frac{%
dX^{I}}{d\alpha }=\{X^{I},\tilde{N}\}~,  \label{N vector}
\end{equation}%
or, more explicitly, 
\begin{equation}
N^{1x}=\frac{d\rho ^{x}}{d\alpha }=0\quad \text{and}\quad N^{2x}=\frac{d\Phi
_{x}}{d\alpha }=1~.  \label{N vector b}
\end{equation}%
The conservation of $\tilde{N}$, eq.(\ref{N conservation}), implies that $%
\tilde{N}$ is the generator of a symmetry, namely, 
\begin{equation}
\frac{d\tilde{H}}{d\alpha }=\{\tilde{H},\tilde{N}\}=0~.
\end{equation}%
Integrating (\ref{N vector b}) one finds the integral curves generated by $%
\tilde{N}$, 
\begin{equation}
\rho ^{x}(\alpha )=\rho ^{x}(0)\quad \text{and}\quad \Phi _{x}(\alpha )=\Phi
_{x}(0)+\alpha ~.
\end{equation}%
This shows that the symmetry generated by $\tilde{N}$ is to shift the phase $%
\Phi $ by a constant $\alpha $ without otherwise changing the dynamics. This
was, of course, already evident in the continuity equation (\ref{CE}) with (%
\ref{drift velocity}) but the implications are very significant. Not only
does the constraint $\tilde{N}=0$ reduce by one the (infinite) number of
independent $\rho ^{x}$ degrees of freedom but the actual number of $\Phi
_{x}$s is also reduced by one because for any value of $\alpha $\ the phases 
$\Phi _{x}+\alpha $\ and $\Phi _{x}$\ correspond to the same state. (This is
the ED analogue of the fact that in QM states are represented by rays rather
than vectors in a Hilbert space.)

The immediate consequence is that two vectors $\bar{U}$ and $\bar{V}$ at $X$
that differ by a vector proportional to $\bar{N}$, 
\begin{equation}
\bar{U}=\bar{V}+k\bar{N}\,,
\end{equation}%
are \textquotedblleft physically\textquotedblright\ equivalent. In
particular the vector $\bar{N}$ is equivalent to zero.

The phase space of interest is $T^{\ast }\mathbf{P}$ but to handle the
constraint $|\rho |{}=1$ we have been led to using coordinates that are more
appropriate to the larger embedding space $T^{\ast }\mathbf{P}^{+1}$. The
price we pay for introducing one superfluous coordinate is to also introduce
a superfluous momentum. We eliminate the extra coordinate by imposing the
constraint $\tilde{N}=0$. We eliminate the extra momentum by declaring it
unphysical. All vectors that differ by a vector along the gauge direction $%
\bar{N}$ are declared equivalent; they belong to the same equivalence class.
The result is a global gauge symmetry.

An equivalence class can be represented by any one of its members and
choosing a convenient representative amounts to fixing the gauge. As we
shall see below a convenient gauge condition is to impose 
\begin{equation}
\int dx\,\rho ^{x}V^{2x}=0\quad \text{or}\quad \langle V^{2}\rangle =0~,
\label{GF vector}
\end{equation}%
so that the representative \textquotedblleft Tangent
Gauge-Fixed\textquotedblright\ vectors (which we shall refer to as TGF
vectors) will satisfy two conditions, eqs.(\ref{tang vector}) and (\ref{GF
vector}), 
\begin{equation}
|V^{1}|{}=\int dx\,V^{1x}=0\quad \text{and}\quad \langle V^{2}\rangle =\int
dx\,\rho ^{x}V^{2x}=0~.  \label{TGF}
\end{equation}%
The first condition enforces a flow tangent to the $\left\vert \rho
\right\vert =1$ surface; the second eliminates a superfluous vector
component along the gauge direction $\bar{N}$.

We end this section with a comment on the symplectic form $\Omega $ which is
non-degenerate on $T^{\ast }\mathbf{P}^{+1}$ but at first sight appears to
be degenerate on $T^{\ast }\mathbf{P}$. Indeed, we have $\Omega (\bar{N},%
\bar{V})=0$ for any tangent vector $\bar{V}$. However, we must recall that $%
\bar{N}$ is equivalent to $0$. In fact, since the TGF equivalent of $\bar{N}$
is $0$, $\Omega $ is not degenerate on $T^{\ast }\mathbf{P}$.

\section{The information geometry of e-phase space}

The construction of the ensemble Hamiltonian $\tilde{H}$ --- or \emph{%
e-Hamiltonian} --- is motivated as follows. The goal of dynamics is to
determine the evolution of the state $(\rho _{t},\Phi _{t})$. From a given
initial state $(\rho _{0},\Phi _{0})$ two slightly different Hamiltonians
will lead to slightly different final states, say $(\rho _{t},\Phi _{t})$ or 
$(\rho _{t}+\delta \rho _{t},\Phi _{t}+\delta \Phi _{t})$. Will these small
changes make any difference? Can we quantify the extent to which we can 
\emph{distinguish} between two neighboring states? This is precisely the
kind of question that metrics are designed to address. It is then natural
that $\tilde{H}$ be in some way related\ to some choice of metric. But
although $\mathbf{P}$ is naturally endowed with a unique information metric
the space $T^{\ast }\mathbf{P}$ has none. Thus, our next goal is to
construct a metric for $T^{\ast }\mathbf{P}$.

Once a metric structure is in place we can ask: does the distance between
two neighboring states --- the extent to which we can \emph{distinguish}
them --- grow, stay the same, or diminish over time? There are many
possibilities here but for pragmatic (and esthetic) reasons we are led to
consider the simplest form of dynamics --- one that preserves the metric.
This leads us to study the Hamilton flows (those that preserve the
symplectic structure) that are also Killing flows (those flows that preserve
the metric structure).

In ED entropic time is constructed so that time (duration) is defined by a
clock provided by the system itself. This leads to require that the
generator $\tilde{H}$ of time translations be defined in terms of the very
same clock that provides the measure of time. Thus, the third and final
ingredient in the construction of $\tilde{H}$ is the requirement is that the
e-Hamiltonian agree with (\ref{Hamiltonian a}) in order to reproduce the
evolution of $\rho $ given by the continuity equation (\ref{Hamilton a}).

In this section our goal is to transform e-phase space $T^{\ast }\mathbf{P}$
from a manifold that is merely symplectic to a manifold that is both
symplectic and Riemannian. The implementation of the other two requirements
on $\tilde{H}$ --- that it generates a Hamilton-Killing flow and that it
agrees with the ED continuity equation ---\ will be tackled in sections \ref%
{HK flows} and \ref{The e-Hamiltonian}.

\subsection{The metric on the embedding space $T^{\ast }\mathbf{P}^{+1}$}

The configuration space $\mathbf{P}$ is a metric space. Our goal here is to
extend its metric --- given by information geometry --- to the full
cotangent bundle, $T^{\ast }\mathbf{P}$. It is convenient to first recall
one derivation of the information metric. In the discrete case the
statistical manifold is the $k$-simplex $\Sigma =\{p=(p^{0}\ldots
p^{k}):\tsum_{i=0}^{k}p^{i}=1\}$. The basic idea is to find the most general
metric consistent with a certain symmetry requirement. To suggest what that
symmetry might be we change to new coordinates $\xi ^{i}=(p^{i})^{1/2}$. In
these new coordinates the equation for the $k$-simplex $\Sigma $ --- the
normalization condition --- reads $\tsum_{i=0}^{k}(\xi ^{i})^{2}=1$ which
suggests the equation of a sphere.

We take this hint seriously and \emph{declare} that the $k$-simplex is a $k$%
-sphere embedded in a generic $(k+1)$-dimensional spherically symmetric
space $\Sigma ^{+1}$.\footnote{%
We are effectively determining the metric by imposing a symmetry, namely,
rotational invariance. One might be concerned that choosing this symmetry is
an ad hoc assumption but the result proves to be very robust. It turns out
that exactly the same metric is obtained by several other criteria that may
appear more natural in the context of inference and probability. Such
criteria include invariance under Markovian embeddings, the geometry of
asymptotic inference, and the metrics induced by relative entropy \cite%
{Campbell 1986}\cite{Rodriguez 1989} (see also \cite{Caticha 2012}).} In the 
$\xi ^{i}$ coordinates the metric of $\Sigma ^{+1}$ is of the form 
\begin{equation}
d\ell ^{2}=\left[ a(|p|)-b(|p|)\right] \left( \tsum_{i=0}^{k}\xi ^{i}d\xi
^{i}\right) ^{2}+|p|{}b(|p|{})\tsum_{i=0}^{k}(d\xi ^{i})^{2},
\label{info metric sph sym a}
\end{equation}%
where $a(|p|)$ and $b(|p|)$ are two arbitrary smooth and positive functions
of $|p|{}=\tsum_{i=0}^{k}p^{i}$. Expressed in terms of the original $p^{i}$
coordinates the metric of $\Sigma ^{+1}$ is 
\begin{equation}
d\ell ^{2}=\left[ a(|p|)-b(|p|)\right] \left( \tsum_{i=0}^{k}dp^{i}\right)
^{2}+|p|b(|p|)\tsum_{i=0}^{k}\frac{1}{p^{i}}(dp^{i})^{2}~.
\label{info metric sph sym b}
\end{equation}%
The restriction to normalized states, $|p|{}=1$ with displacements tangent
to the simplex, $\tsum_{i=0}^{k}dp^{i}=0$,~gives the information metric
induced on the $k$-simplex $\Sigma $,%
\begin{equation}
d\ell ^{2}=b(1)\tsum_{i=0}^{k}\frac{1}{p^{i}}(dp^{i})^{2}~.
\end{equation}%
The overall constant $b(1)$ is not important; it amounts to a choice of the
units of distance.

To extend the information metric from the $k$-simplex $\Sigma $ to its
cotangent bundle $T^{\ast }\Sigma $ we focus on the embedding spaces $\Sigma
^{+1}$ and $T^{\ast }\Sigma ^{+1}$ and require that

\begin{description}
\item[\textbf{(a)}] the metric on $T^{\ast }\Sigma ^{+1}$ be compatible with
the metric on $\Sigma ^{+1}$; and

\item[\textbf{(b)}] that the spherical symmetry of the $(k+1)$-dimensional
space $\Sigma ^{+1}$ be enlarged to full spherical symmetry for the $2(k+1)$%
-dimensional space $T^{\ast }\Sigma ^{+1}$.
\end{description}

\noindent The simplest way to implement \textbf{(a)} is to follow as closely
as possible the derivation that led to (\ref{info metric sph sym b}). The
fact that $\Phi $ inherits from the drift potential $\phi $ the topological
structure of an angle suggests introducing new coordinates, 
\begin{equation}
\xi ^{i}=(p^{i})^{1/2}\cos \Phi _{i}/\hbar \quad \text{and }\quad \eta
^{i}=(p^{i})^{1/2}\sin \Phi _{i}/\hbar ~.
\end{equation}%
Then the normalization condition reads 
\begin{equation}
|p|{}=\tsum_{i=0}^{k}\,p^{i}=\tsum_{i=0}^{k}\left[ (\xi ^{i})^{2}+(\eta
^{i})^{2}\right] =1
\end{equation}%
which suggests the equation of a $(2k+1)$-sphere embedded in $2(k+1)$
dimensions. To implement \textbf{(b)} we take this spherical symmetry
seriously. The most general metric in the embedding space that is invariant
under rotations is 
\begin{eqnarray}
d\ell ^{2} &=&\left[ a(|p|{})-b(|p|{})\right] \left[ \tsum_{i=0}^{k}\left(
\xi ^{i}d\xi ^{i}+\eta ^{i}d\eta ^{i}\right) \right] ^{2}  \notag \\
&&+|p|{}b(|p|{})\tsum_{i=0}^{k}\left[ (d\xi ^{i})^{2}+(d\eta ^{i})^{2}\right]
~,  \label{spherical metric}
\end{eqnarray}%
where the two functions $a(|p|{})$ and $b(|p|{})$ are smooth and positive
but otherwise arbitrary. Therefore, changing back to the $(p^{i},\Phi _{i})$
coordinates, the most general rotationally invariant metric for the
embedding space $T^{\ast }\Sigma ^{+1}$ is 
\begin{eqnarray}
d\ell ^{2} &=&\frac{1}{4}\left[ a(|p|{})-b(|p|{})\right] \left[
\tsum_{i=0}^{k}dp^{i}\right] ^{2}  \notag \\
&&+|p|{}b(|p|{})\frac{1}{2\hbar }\tsum_{i=0}^{k}\left[ \frac{\hbar }{2p^{i}}%
(dp^{i})^{2}+\frac{2p^{i}}{\hbar }(d\Phi _{i})^{2}\right] \,.
\end{eqnarray}

Generalizing from the finite-dimensional case to the $\infty $-dimensional
case yields the metric on the spherically symmetric space $T^{\ast }\mathbf{P%
}^{+1}$, 
\begin{equation}
\delta \tilde{\ell}^{2}=A\left[ \int dx\,\delta \rho _{x}\right] ^{2}+B\int
dx\,\left[ \frac{\hbar }{2\rho _{x}}(\delta \rho _{x})^{2}+\frac{2\rho _{x}}{%
\hbar }(\delta \Phi _{x})^{2}\right] ~.  \label{TP+1 metric}
\end{equation}%
where we set 
\begin{equation}
A(|\rho |)=\frac{1}{4}\left[ a(|\rho |{})-b(|\rho |{})\right] \quad \text{and%
}\quad B(|\rho |)=\frac{1}{2\hbar }|\rho |{}b(|\rho |{})\,.
\end{equation}

\subsection{The metric induced on $T^{\ast }\mathbf{P}$}

As we saw in section \ref{normalization constraint} the normalization
constraint $|\rho |{}=1$ induces a symmetry --- points with phases differing
by a constant are identified. Therefore the e-phase space $T^{\ast }\mathbf{P%
}$ can be obtained from the spherically symmetric space $T^{\ast }\mathbf{P}%
^{+1}$ by the restriction $|\rho |\,=1$ and by identifying points $(\rho
_{x},\Phi _{x})$ and $(\rho _{x},\Phi _{x}+\alpha )$ that lie on the same
gauge orbit, or on the same ray.

Consider two neighboring\ points $(\rho _{x},\Phi _{x})$ and $(\rho
_{x}^{\prime },\Phi _{x}^{\prime })$. The metric induced on $T^{\ast }%
\mathbf{P}$ is defined as the shortest $T^{\ast }\mathbf{P}^{+1}$ distance
between $(\rho _{x},\Phi _{x})$ and points on the ray defined by $(\rho
_{x}^{\prime },\Phi _{x}^{\prime })$. Setting $|\delta \rho |\,=0$ the $%
T^{\ast }\mathbf{P}^{+1}$ distance between $(\rho _{x},\Phi _{x})$ and $%
(\rho _{x}+\delta \rho _{x},\Phi _{x}+\delta \Phi _{x}+\delta \alpha )$ is
given by 
\begin{equation}
\delta \tilde{\ell}^{2}=B(1)\int dx\,\left[ \frac{\hbar }{2\rho _{x}}(\delta
\rho _{x})^{2}+\frac{2\rho _{x}}{\hbar }(\delta \Phi _{x}+\delta \alpha )^{2}%
\right] ~.  \label{TP metric a}
\end{equation}%
Let%
\begin{equation}
\delta \tilde{s}^{2}=\min_{\delta \alpha }\delta \tilde{\ell}^{2}~.
\label{TP metric b}
\end{equation}%
Minimizing over $\delta \alpha $ gives the metric on $T^{\ast }\mathbf{P}$, 
\begin{equation}
\delta \tilde{s}^{2}=\int dx\,\left[ \frac{\hbar }{2\rho _{x}}(\delta \rho
_{x})^{2}+\frac{2\rho _{x}}{\hbar }(\delta \Phi _{x}-\langle \delta \Phi
\rangle )^{2}\right] ~,  \label{TP metric c}
\end{equation}%
where we set $B(1)=1$ which amounts to a choice of units of length. This
metric is known as the Fubini-Study metric.

The scalar product between two vectors $\bar{V}$ and $\bar{U}$ is 
\begin{equation}
G(\bar{V},\bar{U})=\int dx\,\left[ \frac{\hbar }{2\rho _{x}}V^{1x}U^{1x}+%
\frac{2\rho _{x}}{\hbar }(V^{2x}-\langle V^{2}\rangle )(U^{2x}-\langle
U^{2}\rangle )\right] ~.
\end{equation}%
It is at this point that we recognize the convenience of imposing the TGF
gauge condition (\ref{TGF}): the scalar product simplifies to 
\begin{equation}
G(\bar{V},\bar{U})=\int dx\,\left[ \frac{\hbar }{2\rho _{x}}V^{1x}U^{1x}+%
\frac{2\rho _{x}}{\hbar }V^{2x}U^{2x}\right] ~.  \label{ePS a}
\end{equation}%
An analogous expression can be written for the length $\delta \tilde{s}$ of
a displacement $(\delta \rho _{x},\delta \Phi _{x})$, 
\begin{equation}
\delta \tilde{s}^{2}=\int dx\,\left[ \frac{\hbar }{2\rho _{x}}(\delta \rho
_{x})^{2}+\frac{2\rho _{x}}{\hbar }(\delta \Phi _{x})^{2}\right] ~,
\label{ePS b}
\end{equation}%
where it is understood that $(\delta \rho _{x},\delta \Phi _{x})$ satisfies
the TGF condition 
\begin{equation}
|\delta \rho |{}=0\quad \text{and}\quad \langle \delta \Phi \rangle =0~.
\label{TGF b}
\end{equation}
In index notation the metric (\ref{ePS b}) of $T^{\ast }\mathbf{P}$ is
written as 
\begin{equation}
\delta \tilde{s}^{2}=G_{IJ}\delta X^{I}\delta X^{J}=\int dxdx^{\prime
}G_{\alpha x,\beta x^{\prime }}\delta X^{x\alpha }\delta X^{x^{\prime }\beta
}
\end{equation}%
where the metric tensor $G_{IJ}$ is\footnote{%
The tensor $G_{IJ}$\ in eq.(\ref{ePS c}) can act on arbitrary vectors
whether they satisfy the\ TGF condition or not. It is only when $G_{IJ}$
acts on TGF vectors that it is interpreted as a metric tensor on $T^{\ast }%
\mathbf{P}$.} 
\begin{equation}
G_{IJ}=G_{\alpha x,\beta x^{\prime }}=%
\begin{bmatrix}
\frac{\hbar }{2\rho _{x}}\delta _{xx^{\prime }} & 0 \\ 
0 & \frac{2}{\hbar }\rho _{x}\delta _{xx^{\prime }}%
\end{bmatrix}%
~.  \label{ePS c}
\end{equation}%
\noindent

\subsection{A complex structure}

Next we contract the symplectic form $\Omega _{IJ}$, eq.(\ref{sympl form c}%
), with the inverse of the metric tensor, 
\begin{equation}
G^{IJ}=G^{\alpha x,\beta x^{\prime }}=%
\begin{bmatrix}
\frac{2}{\hbar }\rho _{x}\delta _{xx^{\prime }} & 0 \\ 
0 & \frac{\hbar }{2\rho _{x}}\delta _{xx^{\prime }}%
\end{bmatrix}%
~.  \label{ePS c inv}
\end{equation}%
The result is a mixed tensor $J$ with components 
\begin{equation}
J^{I}{}_{J}=-G^{IK}\Omega _{KJ}=%
\begin{bmatrix}
0 & -\frac{2}{\hbar }\rho _{x}\delta _{xx^{\prime }} \\ 
\frac{\hbar }{2\rho _{x}}\delta _{xx^{\prime }} & 0%
\end{bmatrix}%
~.  \label{J tensor}
\end{equation}%
(The reason for introducing an additional negative sign will become clear
below.) The tensor $J^{I}{}_{J}$ maps vectors to vectors --- as any mixed $%
(1,1)$ tensor should. What makes the tensor $J$ special is that --- as one
can easily check --- its action on a TGF vector $\bar{V}$ yields another
vector $J\bar{V}$ that is also TGF and, furthermore, its square is 
\begin{equation}
J^{I}{}_{K}J^{K}{}_{J}=-\delta _{xx^{\prime }}%
\begin{bmatrix}
1 & 0 \\ 
0 & 1%
\end{bmatrix}%
=-\delta ^{I}{}_{J}~.
\end{equation}%
In words, when acting on vectors tangent to $T^{\ast }\mathbf{P}$ the action
of $J^{2}$ (or $\Omega ^{2}$) is equivalent to multiplying by $-1$. This
means that $J$ plays the role of a complex structure.

We conclude that the cotangent bundle $T^{\ast }\mathbf{P}$ has a symplectic
structure $\Omega $, as all cotangent bundles do; that it can be given a
Riemannian structure $G_{IJ}$; and that the mixed tensor $J$ provides it
with a complex structure.

\subsection{Complex coordinates}

The fact that $T^{\ast }\mathbf{P}$ is endowed with a complex structure
suggests introducing complex coordinates, 
\begin{equation}
\Psi _{x}=\rho _{x}^{1/2}\exp i\Phi _{x}/\hbar ~,  \label{Psi}
\end{equation}%
so that a point $\Psi \in $ $T^{\ast }\mathbf{P}^{+1}$ has coordinates 
\begin{equation}
\Psi ^{\mu x}=\binom{\Psi ^{1x}}{\Psi ^{2x}}=\binom{\Psi _{x}}{i\hbar \Psi
_{x}^{\ast }}~,
\end{equation}%
where the index $\mu $ takes two values, $\mu =1,2$.

We can check that the transformation from real coordinates $(\rho ,\Phi )$
to complex coordinates $(\Psi ,i\hbar \Psi ^{\ast })$ is canonical. Indeed,
the action of $\Omega $ on two infinitesimal vectors $\delta X^{I}$ and $%
\delta ^{\prime }X^{J}$ is 
\begin{equation*}
\Omega _{IJ}\delta X^{I}\delta ^{\prime }X^{J}=\int dx\,\left( \delta \rho
_{x}\delta ^{\prime }\Phi _{x}-\delta \Phi _{x}\delta ^{\prime }\rho
_{x}\right) ~,
\end{equation*}%
which, when expressed in $\Psi $ coordinates, becomes 
\begin{equation}
\Omega _{IJ}\delta X^{I}\delta ^{\prime }X^{J}=\int dx\,\left( \delta \Psi
\delta ^{\prime }i\hbar \Psi ^{\ast }-\delta i\hbar \Psi ^{\ast }\delta
^{\prime }\Psi \right) =\Omega _{\mu x,\nu x^{\prime }}\delta \Psi ^{\mu
x}\delta \Psi ^{\nu x^{\prime }}
\end{equation}%
where 
\begin{equation}
\Omega _{\mu x,\nu x^{\prime }}=%
\begin{bmatrix}
0 & 1 \\ 
-1 & 0%
\end{bmatrix}%
\delta _{xx^{\prime }}~,  \label{sympl form d}
\end{equation}%
retains the same form as (\ref{sympl form c}).

Expressed in $\Psi $ coordinates the Hamiltonian flow generated by the
normalization constraint (\ref{N constraint}), 
\begin{equation}
\tilde{N}=0\quad \text{with}\quad \tilde{N}=1-\int dx\,\Psi _{x}^{\ast }\Psi
_{x}~,
\end{equation}%
and parametrized by $\alpha $ is given by the vector field 
\begin{equation}
\bar{N}=-\binom{\Psi _{x}/i\hbar }{i\hbar (\Psi _{x}/i\hbar )^{\ast }}~.
\end{equation}%
Its integral curves are 
\begin{equation}
\Psi _{x}(\alpha )=\Psi _{x}(0)e^{i\alpha /\hbar }~.
\end{equation}%
The constraint $\tilde{N}=0$ induces a gauge symmetry which leads us to
restrict our attention to vectors $\bar{V}=d/d\lambda $ satisfying two real
TGF conditions (\ref{TGF}). In $\Psi $ coordinates this is replaced by the
single complex TGF condition, 
\begin{equation}
\int dx\,\Psi _{x}^{\ast }\frac{d\Psi _{x}}{d\lambda }=0~.  \label{TGF Psi}
\end{equation}

In $\Psi $ coordinates the metric on $T^{\ast }\mathbf{P}$, eq.(\ref{ePS b}%
), becomes 
\begin{equation}
\delta \tilde{s}^{2}=-2i\int dx\,\delta \Psi _{x}\delta i\hbar \Psi
_{x}^{\ast }=\int dxdx^{\prime }G_{\mu x,\nu x^{\prime }}\,\delta \Psi ^{\mu
x}\delta \Psi ^{\nu x^{\prime }}~,  \label{metric Psi a}
\end{equation}%
where the metric tensor and its inverse are 
\begin{equation}
G_{\mu x,\nu x^{\prime }}=-i\delta _{xx^{\prime }}%
\begin{bmatrix}
0 & 1 \\ 
1 & 0%
\end{bmatrix}%
\quad \text{and}\quad G^{\mu x,\nu x^{\prime }}=i\delta _{xx^{\prime }}%
\begin{bmatrix}
0 & 1 \\ 
1 & 0%
\end{bmatrix}%
~.  \label{metric Psi b}
\end{equation}%
Finally, using $G^{\mu x,\nu x^{\prime }}$ to raise the first index of $%
\Omega _{\nu x^{\prime },\gamma x^{\prime \prime }}$ gives the $\Psi $
components of the tensor $J$ 
\begin{equation}
J^{\mu x}{}_{\gamma x^{\prime \prime }}\overset{\text{def}}{=}-G^{\mu x,\nu
x^{\prime }}\Omega _{\nu x^{\prime },\gamma x^{\prime \prime }}=%
\begin{bmatrix}
i & 0 \\ 
0 & -i%
\end{bmatrix}%
\delta _{xx^{\prime }}~.  \label{complex structure J}
\end{equation}

\section{Hamilton-Killing flows}

\label{HK flows}Our next goal will be to find those Hamiltonian flows $Q^{I}$
that also happen to preserve the metric tensor, that is, we want $Q^{I}$ to
be a Killing vector. The condition for $Q^{I}$ is 
\begin{equation}
(\pounds _{Q}G)_{IJ}=Q^{K}\tilde{\nabla}_{K}G_{IJ}+G_{KJ}\tilde{\nabla}%
_{I}Q^{K}+G_{IK}\tilde{\nabla}_{J}Q^{K}=0~.
\end{equation}%
In complex coordinates eq.(\ref{metric Psi b}) gives $\tilde{\nabla}%
_{K}G_{IJ}=0$, and the Killing equation simplifies to 
\begin{equation}
(\pounds _{Q}G)_{IJ}=G_{KJ}\tilde{\nabla}_{I}Q^{K}+G_{IK}\tilde{\nabla}%
_{J}Q^{K}=0~,
\end{equation}%
or 
\begin{equation}
(\pounds _{Q}G)_{\mu x,\nu x^{\prime }}=-i%
\begin{bmatrix}
\frac{\delta Q^{2x^{\prime }}}{\delta \Psi _{x}}+\frac{\delta Q^{2x}}{\delta
\Psi _{x^{\prime }}}~; & \frac{\delta Q^{1x^{\prime }}}{\delta \Psi _{x}}+%
\frac{\delta Q^{2x}}{\delta i\hbar \Psi _{x^{\prime }}^{\ast }} \\ 
\frac{\delta Q^{2x^{\prime }}}{\delta i\hbar \Psi _{x}^{\ast }}+\frac{\delta
Q^{1x}}{\delta \Psi _{x^{\prime }}}~; & \frac{\delta Q^{1x^{\prime }}}{%
\delta i\hbar \Psi _{x}^{\ast }}+\frac{\delta Q^{1x}}{\delta i\hbar \Psi
_{x^{\prime }}^{\ast }}%
\end{bmatrix}%
=0~.  \label{K flow a}
\end{equation}%
If we further require that $Q^{I}$ be a Hamiltonian flow, $\pounds %
_{Q}\Omega =0$, then we substitute 
\begin{equation}
Q^{1x}=\frac{\delta \tilde{Q}}{\delta i\hbar \Psi _{x}^{\ast }}\quad \text{%
and}\quad Q^{2x}=-\frac{\delta \tilde{Q}}{\delta \Psi _{x}}~
\label{H flow a}
\end{equation}%
into (\ref{K flow a}) to get 
\begin{equation}
\frac{\delta ^{2}\tilde{Q}}{\delta \Psi _{x}\delta \Psi _{x^{\prime }}}%
=0\quad \text{and}\quad \frac{\delta ^{2}\tilde{Q}}{\delta \Psi _{x}^{\ast
}\delta \Psi _{x^{\prime }}^{\ast }}=0~.
\end{equation}%
Therefore in order to generate a flow that preserves both $G$ and $\Omega $
the functional $\tilde{Q}[\Psi ,\Psi ^{\ast }]$ must be \emph{linear} in
both $\Psi $ and $\Psi ^{\ast }$, 
\begin{equation}
\tilde{Q}[\Psi ,\Psi ^{\ast }]=\int dxdx^{\prime }\,\Psi _{x}^{\ast }\hat{Q}%
_{xx^{\prime }}\Psi _{x^{\prime }}~,  \label{bilinear hamiltonian}
\end{equation}%
where $\hat{Q}_{xx^{\prime }}$ is a possibly non-local kernel. The actual
Hamilton-Killing flow is 
\begin{eqnarray}
\frac{d\Psi _{x}}{d\lambda } &=&Q^{1x}=\frac{\delta \tilde{Q}}{\delta i\hbar
\Psi _{x}^{\ast }}=\frac{1}{i\hbar }\int dx^{\prime }\,\hat{Q}_{xx^{\prime
}}\Psi _{x^{\prime }}~,  \label{HK flow 1} \\
\frac{di\hbar \Psi _{x}^{\ast }}{d\lambda } &=&Q^{2x}=-\frac{\delta \tilde{Q}%
}{\delta \Psi _{x}}=-\int dx^{\prime }\,\Psi _{x^{\prime }}^{\ast }\hat{Q}%
_{xx^{\prime }}~.  \label{HK flow 2}
\end{eqnarray}%
Taking the complex conjugate of (\ref{HK flow 1}) and comparing with (\ref%
{HK flow 2}), shows that the kernel $\hat{Q}_{xx^{\prime }}$ is Hermitian, 
\begin{equation}
\hat{Q}_{xx^{\prime }}^{\ast }=\hat{Q}_{x^{\prime }x}~,
\label{V selfadjoint}
\end{equation}%
and we can check that the corresponding Hamiltonian functionals $\tilde{Q}$
are real, 
\begin{equation*}
\tilde{Q}[\Psi ,\Psi ^{\ast }]^{\ast }=\tilde{Q}[\Psi ,\Psi ^{\ast }]~.
\end{equation*}

The Hamiltonian flows that might potentially be of interest are those that
generate symmetry transformations. For example, the generator of
translations is total momentum. Under a spatial displacement by $\varepsilon
^{a}$, $g(x)\rightarrow g_{\varepsilon }(x)=g(x-\varepsilon )$, the change
in $f[\rho ,\Phi ]$ is 
\begin{equation}
\delta _{\varepsilon }f[\rho ,\Phi ]=\int dx\left( \frac{\delta f}{\delta
\rho _{x}}\delta _{\varepsilon }\rho _{x}+\frac{\delta f}{\delta \Phi _{x}}%
\delta _{\varepsilon }\Phi _{x}\right) =\{f,\tilde{P}_{a}\varepsilon ^{a}\}
\end{equation}%
where 
\begin{equation}
\tilde{P}_{a}=\int dx\,\rho \dsum\nolimits_{n}\frac{\partial \Phi }{\partial
x_{n}^{a}}=\int dx\,\rho \frac{\partial \Phi }{\partial X^{a}}
\end{equation}%
is interpreted as the expectation of the total momentum, and $X^{a}$ are the
coordinates of the center of mass, 
\begin{equation}
X^{a}=\frac{1}{M}\dsum\nolimits_{n}m_{n}x_{n}^{a}~.  \label{CM}
\end{equation}%
In complex coordinates, 
\begin{equation}
\tilde{P}_{a}=\int dx\,\Psi ^{\ast }\left( \dsum\nolimits_{n}\frac{\hbar }{i}%
\frac{\partial }{\partial x_{n}^{a}}\right) \Psi =\int dx\,\Psi ^{\ast
}\left( \frac{\hbar }{i}\frac{\partial }{\partial X^{a}}\right) \Psi \ ,
\end{equation}%
and the corresponding kernel $\hat{P}_{axx^{\prime }}$ is 
\begin{equation}
\hat{P}_{axx^{\prime }}=\delta _{xx^{\prime }}\dsum\nolimits_{n}\frac{\hbar 
}{i}\frac{\partial }{\partial x_{n}^{a}}=\delta _{xx^{\prime }}\frac{\hbar }{%
i}\frac{\partial }{\partial X^{a}}~.
\end{equation}

\section{The e-Hamiltonian}

\label{The e-Hamiltonian}In previous sections we supplied the symplectic
e-phase space $T^{\ast }\mathbf{P}$ with a Riemannian metric and, as a
welcome by-product, also with a complex structure. Then we showed that the
condition for the simplest form of dynamics --- one that preserves all the
metric, symplectic, and complex structures --- is a Hamilton-Killing flow
generated by a Hamiltonian $\tilde{H}$ that is linear in both $\Psi $ and $%
\Psi ^{\ast }$, 
\begin{equation}
\tilde{H}[\Psi ,\Psi ^{\ast }]=\int dxdx^{\prime }\,\Psi _{x}^{\ast }\hat{H}%
_{xx^{\prime }}\Psi _{x^{\prime }}~.
\end{equation}%
The last ingredient in the construction of $\tilde{H}$ is that the
e-Hamiltonian has to agree with (\ref{Hamiltonian a}) in order to reproduce
the entropic evolution of $\rho $ given by the continuity eq.(\ref{Hamilton
a}).

To proceed we use the identity 
\begin{equation}
\frac{1}{2}\rho m^{AB}(\partial _{A}\Phi -\bar{A}_{A})(\partial _{B}\Phi -%
\bar{A}_{B})=\frac{\hbar ^{2}}{2}m^{AB}(D_{A}\Psi )^{\ast }D_{B}\Psi -\frac{%
\hbar ^{2}}{8\rho ^{2}}m^{AB}\partial _{A}\rho \partial _{B}\rho
\end{equation}%
where 
\begin{equation}
D_{A}=\partial _{A}-\frac{i}{\hbar }\bar{A}_{A}\quad \text{and}\quad \bar{A}%
_{A}(x)=\hbar \beta _{n}A_{a}(x_{n})~.
\end{equation}%
Rewriting $\tilde{H}[\rho ,\Phi ]$ in (\ref{Hamiltonian a}) in terms of $%
\Psi $ and $\Psi ^{\ast }$ we get 
\begin{equation}
\tilde{H}[\Psi ,\Psi ^{\ast }]=\int dx\left( \frac{-\hbar ^{2}}{2}m^{AB}\Psi
^{\ast }D_{A}D_{B}\Psi \right) +F^{\prime }[\rho ]~.
\end{equation}%
where 
\begin{equation}
F^{\prime }[\rho ]=F[\rho ]-\frac{\hbar ^{2}}{8\rho ^{2}}m^{AB}\partial
_{A}\rho \partial _{B}\rho ~.  \label{F prime a}
\end{equation}%
According to (\ref{bilinear hamiltonian}) in order for $\tilde{H}[\Psi ,\Psi
^{\ast }]$ to generate an HK flow we must impose that $F^{\prime }[\rho ]$
be linear in both $\Psi $ and $\Psi ^{\ast }$, 
\begin{equation}
F^{\prime }[\rho ]=\int dxdx^{\prime }\,\Psi _{x}^{\ast }\hat{V}_{xx^{\prime
}}\Psi _{x^{\prime }}  \label{F prime b}
\end{equation}%
for some Hermitian kernel $\hat{V}_{xx^{\prime }}$, but $F^{\prime }[\rho ]$
must remain independent of $\Phi $, 
\begin{equation}
\frac{\delta F^{\prime }[\rho ]}{\delta \Phi _{x}}=0~.  \label{F prime c}
\end{equation}%
Substituting $\Psi =\rho ^{1/2}e^{i\Phi /\hbar }$ into (\ref{F prime b}) and
using $\hat{V}_{x^{\prime }x}^{\ast }=\hat{V}_{xx^{\prime }}$ leads to%
\begin{equation}
\frac{\delta F^{\prime }}{\delta \Phi _{x}}=\frac{2}{\hbar }\rho
_{x}^{1/2}\int dx^{\prime }\rho _{x^{\prime }}^{1/2}\func{Im}\left( \hat{V}%
_{xx^{\prime }}e^{-i\left( \Phi _{x}-\Phi _{x^{\prime }}\right) /\hbar
}\right) =0
\end{equation}%
This equation must be satisfied for all choices of $\rho _{x^{\prime }}$,
which implies 
\begin{equation}
\func{Im}\left( \hat{V}_{xx^{\prime }}e^{-i\left( \Phi _{x}-\Phi _{x^{\prime
}}\right) /\hbar }\right) =0~,
\end{equation}%
and also for all choices of $\Phi _{x}$ and $\Phi _{x^{\prime }}$.
Therefore, the kernel $\hat{V}_{xx^{\prime }}$ must be local in $x$, 
\begin{equation}
\hat{V}_{xx^{\prime }}=\delta _{xx^{\prime }}V_{x}~,
\end{equation}%
where $V_{x}=V(x)$ is some real function.

We conclude that the Hamiltonian that generates a Hamilton-Killing flow and
agrees with the ED continuity equation must be of the form 
\begin{equation}
\tilde{H}[\Psi ,\Psi ^{\ast }]=\int dx\Psi ^{\ast }\left( -\frac{\hbar ^{2}}{%
2}m^{AB}D_{A}D_{B}+V(x)\right) \Psi ~.  \label{Hamiltonian c}
\end{equation}

The evolution of $\Psi $ is given by the Hamilton equation, 
\begin{equation}
\partial _{t}\Psi _{x}=\{\Psi _{x},\tilde{H}\}=\frac{\delta \tilde{H}}{%
\delta (i\hbar \Psi ^{\ast }(x))}~,
\end{equation}%
which is the Schr\"{o}dinger equation, 
\begin{equation}
i\hbar \partial _{t}\Psi =-\frac{\hbar ^{2}}{2}m^{AB}D_{A}D_{B}\Psi +V\Psi ~.
\label{sch a}
\end{equation}%
In more standard notation it reads 
\begin{equation}
i\hbar \partial _{t}\Psi =\dsum\nolimits_{n}\frac{-\hbar ^{2}}{2m_{n}}\delta
^{ab}\left( \frac{\partial }{\partial x_{n}^{a}}-i\beta
_{n}A_{a}(x_{n})\right) \left( \frac{\partial }{\partial x_{n}^{b}}-i\beta
_{n}A_{b}(x_{n})\right) \Psi +V\Psi ~.  \label{sch b}
\end{equation}

At this point we can finally provide the physical interpretation of the
various constants introduced along the way. Since the Schr\"{o}dinger
equation (\ref{sch b}) is the tool we use to analyze experimental data we
can identify $\hbar $ with Planck's constant, $m_{n}$ will be interpreted as
the particles' masses, and the $\beta _{n}$ are related to the particles'
electric charges $q_{n}$ by%
\begin{equation}
\beta _{n}=\frac{q_{n}}{\hbar c}~.  \label{charge a}
\end{equation}

For completeness we write the Hamiltonian in the $(\rho ,\Phi )$ variables, 
\begin{eqnarray}
\tilde{H}[\rho ,\Phi ] &=&\int d^{3N}x\,\rho \left[ \dsum\nolimits_{n}\frac{%
\delta ^{ab}}{2m_{n}}\left( \frac{\partial \Phi }{\partial x_{n}^{a}}-\frac{%
q_{n}}{c}A_{a}(x_{n})\right) \left( \frac{\partial \Phi }{\partial x_{n}^{b}}%
-\frac{q_{n}}{c}A_{b}(x_{n})\right) \right.  \notag \\
&&+\left. \dsum\nolimits_{n}\frac{\hbar ^{2}}{8m_{n}}\frac{\delta ^{ab}}{%
\rho ^{2}}\frac{\partial \rho }{\partial x_{n}^{a}}\frac{\partial \rho }{%
\partial x_{n}^{b}}+V(x_{1}\ldots x_{n})\right] ~.
\end{eqnarray}%
The Hamilton equations for $\rho $ and $\Phi $ are the continuity equation (%
\ref{Hamilton a}), 
\begin{equation}
\partial _{t}\rho =\frac{\delta \tilde{H}}{\delta \Phi }=-\dsum\nolimits_{n}%
\frac{\partial }{\partial x_{n}^{a}}\left[ \rho \frac{\delta ^{ab}}{m_{n}}%
\left( \frac{\partial \Phi }{\partial x_{n}^{b}}-\frac{q_{n}}{c}%
A_{b}(x_{n})\right) \right] ~,  \label{CE c}
\end{equation}%
and the quantum analogue of the Hamilton-Jacobi equation, \ 
\begin{eqnarray}
\partial _{t}\Phi &=&-\frac{\delta \tilde{H}}{\delta \rho }%
=\dsum\nolimits_{n}\frac{-\delta ^{ab}}{2m_{n}}\left( \frac{\partial \Phi }{%
\partial x_{n}^{a}}-\frac{q_{n}}{c}A_{a}(x_{n})\right) \left( \frac{\partial
\Phi }{\partial x_{n}^{b}}-\frac{q_{n}}{c}A_{b}(x_{n})\right)  \notag \\
&&+\left. \dsum\nolimits_{n}\frac{\hbar ^{2}}{2m_{n}}\frac{\delta ^{ab}}{%
\rho ^{1/2}}\frac{\partial ^{2}\rho ^{1/2}}{\partial x_{n}^{a}\partial
x_{n}^{b}}-V(x_{1}\ldots x_{n})\right] \,.  \label{HJ}
\end{eqnarray}

To summarize: we have just shown that an ED that preserves both the
symplectic and metric structures of the e-phase space $T^{\ast }\mathbf{P}$
leads to a linear Schr\"{o}dinger equation. In particular, such an ED
reproduces the quantum potential in (\ref{HJ}) with the correct coefficients 
$\hbar ^{2}/2m_{n}$.

\section{Entropic time, physical time, and time reversal}

\label{entropic v physical time}Now that the dynamics has been fully
developed we revisit the question of time. The derivation of laws of physics
as examples of inference led us to introduce the notion of entropic time
which includes assumptions about the concept of instant, of simultaneity, of
ordering, and of duration. It is clear that entropic time is useful but is
this the actual, real, \textquotedblleft physical\textquotedblright\ time?
The answer is yes. By deriving the Schr\"{o}dinger equation (from which we
can obtain the classical limit) we have shown that the $t$ that appears in
the laws of physics is entropic time. Since these are the equations that we
routinely use to design and calibrate our clocks we conclude that \emph{what
clocks measure is entropic time}. No notion of time that is in any way
deeper or more \textquotedblleft physical\textquotedblright\ is needed. Most
interestingly, the entropic model automatically includes an arrow of time.

The statement that the laws of physics are invariant under time reversal has
nothing to do with particles travelling backwards in time. It is instead the
assertion that the laws of physics exhibit a certain symmetry. For a
classical system described by coordinates $q$ and momenta $p$ the symmetry
is the statement that if $\{q_{t},p_{t}\}$ happens to be one solution of
Hamilton's equations then we can construct another solution $%
\{q_{t}^{T},p_{t}^{T}\}$ where 
\begin{equation}
q_{t}^{T}=q_{-t}\quad \text{and}\quad p_{t}^{T}=-p_{-t}~,
\end{equation}%
but both solutions $\{q_{t},p_{t}\}$ and $\{q_{t}^{T},p_{t}^{T}\}$ describe
evolution forward in time. An alternative statement of time reversibility is
the following\textbf{:} if there is one trajectory of the system that takes
it from state $\{q_{0},p_{0}\}$ at time $t_{0}$ to state $\{q_{1},p_{1}\}$
at the later time $t_{1}$, then there is another possible trajectory that
takes the system from state $\{q_{1},-p_{1}\}$ at time $t_{0}$ to state $%
\{q_{0},-p_{0}\}$ at the later time $t_{1}$. The merit of this re-statement
is that it makes clear that nothing needs to travel back in time. Indeed,
rather than time reversal the symmetry might be more appropriately described
as momentum or motion reversal.

Since ED is a Hamiltonian dynamics one can expect that similar
considerations will apply to QM and indeed they do. It is straightforward to
check that given one solution $\{\rho _{t}(x),\Phi _{t}(x)\}$ that evolves
forward in time, we can construct another solution $\{\rho _{t}^{T}(x),\Phi
_{t}^{T}(x)\}$ that is also evolving forward in time. The reversed solution
is 
\begin{equation}
\rho _{t}^{T}(x)=\rho _{-t}(x)\quad \text{and}\quad \Phi _{t}^{T}(x)=-\Phi
_{-t}(x)~.
\end{equation}%
These transformations constitute a symmetry --- \emph{i.e.}, the transformed 
$\Psi _{t}^{T}(x)$ is a solution of the Schr\"{o}dinger equation ---
provided the motion of the sources of the external potentials is also
reversed, that is, the potentials $A_{a}(\vec{x},t)$ and $V(x,t)$ are
transformed according to 
\begin{equation}
A_{a}^{T}(\vec{x},t)=-A_{a}(\vec{x},-t)\quad \text{and}\quad
V^{T}(x,t)=V(x,-t)~.
\end{equation}%
Expressed in terms of wave functions the time reversal transformation is 
\begin{equation}
\Psi _{t}^{T}(x)=\Psi _{-t}^{\ast }(x)~.
\end{equation}%
The proof that this is a symmetry is straightforward; just take the complex
conjugate of (\ref{sch b}), and let $t\rightarrow -t$.

\section{Linearity and the superposition principle}

The Schr\"{o}dinger equation is linear, that is, a linear combination of
solutions is a solution too. However, this \emph{mathematical} linearity
does not guarantee the \emph{physical} linearity that is usually referred to
as the superposition principle. The latter is the physical assumption that
if there is one experimental setup that prepares a system in the (epistemic)
state $\Psi _{1}$ and there is another setup that prepares the system in the
state $\Psi _{2}$ then, at least in principle, it is possible to construct
yet a third setup that can prepare the system in the superposition 
\begin{equation}
\Psi _{3}=\alpha _{1}\Psi _{1}+\alpha _{2}\Psi _{2}~,  \label{psi3}
\end{equation}%
where $\alpha _{1}$ and $\alpha _{2}$ are arbitrary complex numbers.
Mathematical linearity refers to the fact that solutions can be expressed as
sums of solutions. There is no implication that any of these solutions will
necessarily describe physical situations. Physical linearity on the other
hand --- the Superposition Principle --- refers to the fact that the
superposition of physical solutions is also a physical solution. The point
to be emphasized is the Superposition Principle is not a principle; it is a
physical hypothesis that need not be universally true.

\subsection{The single-valuedness of $\Psi $}

The question \textquotedblleft Why should wave functions be
single-valued?\textquotedblright\ has been around for a long time. In this
section we build on and extend recent work \cite{Carrara Caticha 2017} to
argue that the single- or multi-valuedness of the wave functions is closely
related to the question of linearity and the superposition principle. Our
discussion parallels that by Schr\"{o}dinger \cite{Schrodinger 1938}.%
\footnote{%
Schr\"{o}dinger invoked time reversal invariance which was a very legitimate
move back in 1938 but today it is preferable to develop an argument which
does not invoke symmetries that are already known to be violated.} (See also 
\cite{Pauli 1939}-\cite{Wallstrom 1994}.)\footnote{%
The answer proposed by Pauli is also worthy of note \cite{Pauli 1939}-\cite%
{Merzbacher 1962}. He proposed that admissible wave functions must form a
basis for representations of the transformation group that happens to be
pertinent to the problem at hand. In particular, Pauli's argument serves to
discard double-valued wave functions for describing the orbital angular
momentum of scalar particles. The question of single-valuedness was revived
by Takabayashi \cite{Takabayasi 1952} in the context of the hydrodynamical
interpretation of QM, and later rephrased by Wallstrom \cite{Wallstrom 1989}%
\cite{Wallstrom 1994} as an objection to Nelson's stochastic mechanics: are
these theories equivalent to QM or do they merely reproduce a subset of its
solutions? Wallstrom's objection is that Nelson's stochastic mechanics leads
to phases and wave functions that are either both multi-valued or both
single-valued. Both alternatives are unsatisfactory because on one hand QM
requires single-valued wave functions, while on the other hand single-valued
phases exclude states that are physically relevant (\emph{e.g.}, states with
non-zero angular momentum).}

To show that the mathematical linearity of (\ref{sch b}) is not sufficient
to imply the superposition principle we argue that even when $|\Psi
_{1}|^{2}=\rho _{1}$ and $|\Psi _{2}|^{2}=\rho _{2}$ are probabilities it is
not generally true that $|\Psi _{3}|^{2}$, eq.(\ref{psi3}), will also be a
probability. Consider moving around a closed loop $\Gamma $ in configuration
space. Since phases $\Phi (x)$ can be multi-valued the corresponding wave
functions could in principle be multi-valued too. Let a generic $\Psi $
change by a phase factor, 
\begin{equation}
\Psi \rightarrow \Psi ^{\prime }=e^{i\delta }\Psi ~,  \label{psi delta}
\end{equation}%
then the superposition $\Psi _{3}$ of two wave functions $\Psi _{1}$ and $%
\Psi _{2}$ changes into 
\begin{equation}
\Psi _{3}\rightarrow \Psi _{3}^{\prime }=\alpha _{1}e^{i\delta _{1}}\Psi
_{1}+\alpha _{2}e^{i\delta _{2}}\Psi _{2}~.
\end{equation}%
The problem is that even if $|\Psi _{1}|^{2}=\rho _{1}$ and $|\Psi
_{2}|^{2}=\rho _{2}$ are single-valued (because they are probability
densities), the quantity $|\Psi _{3}|^{2}$\ need not in general be
single-valued. Indeed, 
\begin{equation}
|\Psi _{3}|^{2}=|\alpha _{1}|^{2}\rho _{1}+|\alpha _{2}|^{2}\rho _{2}+2\func{%
Re}[\alpha _{1}\alpha _{2}^{\ast }\Psi _{1}\Psi _{2}^{\ast }]~,
\end{equation}%
changes into 
\begin{equation}
|\Psi _{3}^{\prime }|^{2}=|\alpha _{1}|^{2}\rho _{1}+|\alpha _{2}|^{2}\rho
_{2}+2\func{Re}[\alpha _{1}\alpha _{2}^{\ast }e^{i(\delta _{1}-\delta
_{2})}\Psi _{1}\Psi _{2}^{\ast }]~,
\end{equation}%
so that in general 
\begin{equation}
|\Psi _{3}^{\prime }|^{2}\neq |\Psi _{3}|^{2}~,
\end{equation}%
which precludes the interpretation of $|\Psi _{3}|^{2}$ as a probability.
That is, even when the epistemic states $\Psi _{1}$ and $\Psi _{2}$ describe
actual physical situations, their superpositions need not.

The problem does not arise when 
\begin{equation}
e^{i(\delta _{1}-\delta _{2})}=1~.  \label{SS a}
\end{equation}%
If we were to group the wave functions into classes each characterized by
its own $\delta $ then we could have a limited version of the superposition
principle that applies within each class. We conclude that beyond the
linearity of the Schr\"{o}dinger equation we have a superselection rule that
restricts the validity of the superposition principle to wave functions
belong to the same $\delta $-class.

To find the allowed values of $\delta $ we argue as follows. It is natural
to assume that if $\{\rho ,\Phi \}$ (at some given time $t_{0}$) is a
physical state then the state with reversed momentum $\{\rho ,-\Phi \}$ (at
the same time $t_{0}$) is an equally reasonable physical state.\ Basically,
the idea is that if particles can be prepared to move in one direction, then
they can also be prepared to move in the opposite direction. In terms of
wave functions the statement is that if $\Psi _{t_{0}}$ is a physically
allowed initial state, then so is $\Psi _{t_{0}}^{\ast }$.\footnote{%
We make no symmetry assumptions such as parity or time reversibility. It
need not be the case that there is any symmetry that relates the time
evolution of $\Psi _{t_{0}}^{\ast }$ to that of $\Psi _{t_{0}}$.} Next we
consider a generic superposition 
\begin{equation}
\Psi _{3}=\alpha _{1}\Psi +\alpha _{2}\Psi ^{\ast }~.  \label{SS b}
\end{equation}%
Is it physically possible to construct superpositions such as (\ref{SS b})?
The answer is that while constructing $\Psi _{3}$ for an arbitrary $\Psi $
might not be feasible in practice there is strong empirical evidence that
there exist no superselection rules to prevent us from doing so in
principle. Indeed, it is easy to construct superpositions of wavepackets
with momentum $\vec{p}$ and $-\vec{p}$, or superpositions of states with
opposite angular momenta, $Y_{\ell m}$ and $Y_{\ell ,-m}$. \emph{We shall
assume that in principle the superpositions (\ref{SS b}) are physically
possible.}

According to eq.(\ref{psi delta}) as one moves in a closed loop $\Gamma $
the wave function $\Psi _{3}$ will transform into 
\begin{equation}
\Psi _{3}^{\prime }=\alpha _{1}e^{i\delta }\Psi +\alpha _{2}e^{-i\delta
}\Psi ^{\ast }~,
\end{equation}%
and the condition (\ref{SS a}) for $|\Psi _{3}|^{2}$ to be single-valued is 
\begin{equation}
e^{2i\delta }=1\quad \text{or}\quad e^{i\delta }=\pm 1~.  \label{SS c}
\end{equation}%
Thus, we are restricted to two discrete possibilities $\pm 1$. Since the
wave functions are assumed sufficiently well behaved (continuous,
differentiable, etc.) we conclude that they must be either single-valued, $%
e^{i\delta }=1$, or double-valued, $e^{i\delta }=-1$.

We conclude that the Superposition Principle appears to be valid in a
sufficiently large number of cases to be a useful rule of thumb but it is
restricted to single-valued (or double-valued) wave functions. The argument
above does not exclude the possibility that a multi-valued wave function
might describe an actual physical situation. What the argument implies is
that the Superposition Principle would not extend to such states. \emph{\ }

\subsection{Charge quantization}

Next we analyze the conditions for the electromagnetic gauge symmetry to be
compatible with the superposition principle. We shall confine our attention
to systems that are described by single-valued wave functions ($e^{i\delta
}=+1$).\footnote{%
Double-valued wave functions with $e^{i\delta }=-1$ will, of course, find
use in the description of spin-1/2 particles. \cite{Caticha Carrara 2019}}
The condition for the wave function to be single-valued is%
\begin{equation}
\Delta \frac{\Phi }{\hbar }=\doint_{\Gamma }d\ell ^{A}\partial _{A}\frac{%
\Phi }{\hbar }=2\pi k_{\Gamma }~,  \label{circ}
\end{equation}%
where $k_{\Gamma }$ is an integer that depends on the loop $\Gamma $. Under
a local gauge transformation 
\begin{equation}
A_{a}(\vec{x})\rightarrow A_{a}(\vec{x})+\partial _{a}\chi (\vec{x})
\label{GT 1}
\end{equation}%
the phase $\Phi $ transforms according to (\ref{GT phase}), 
\begin{equation}
\Phi (x)\rightarrow \Phi ^{\prime }(x)=\Phi (x)+\tsum\nolimits_{n}\frac{q_{n}%
}{c}\chi (\vec{x}_{n})~.  \label{GT 2}
\end{equation}%
The requirement that the gauge symmetry and the superposition principle be
compatible amounts to requiring that the gauge transformed states also be
single-valued, 
\begin{equation}
\Delta \frac{\Phi ^{\prime }}{\hbar }=\doint_{\Gamma }d\ell ^{A}\partial _{A}%
\frac{\Phi ^{\prime }}{\hbar }=2\pi k_{\Gamma }^{\prime }~.
\label{circ prime}
\end{equation}%
Thus, the allowed gauge transformations are restricted to functions $\chi (%
\vec{x})$ such that 
\begin{equation}
\tsum\nolimits_{n}\frac{q_{n}}{\hbar c}\doint_{\Gamma }d\ell
_{n}^{a}\partial _{na}\chi (\vec{x}_{n})=2\pi \Delta k_{\Gamma }
\end{equation}%
where $\Delta k_{\Gamma }=k_{\Gamma }^{\prime }-k_{\Gamma }$ is an integer.
Consider now a loop $\gamma $ in which we follow the coordinates of the $n$%
th particle around some closed path in 3-dimensional space while all the
other particles are kept fixed. Then 
\begin{equation}
\frac{q_{n}}{\hbar c}\doint_{\gamma }d\ell _{n}^{a}\partial _{an}\chi (\vec{x%
}_{n})=2\pi \Delta k_{n\gamma }  \label{circ b}
\end{equation}%
where $\Delta k_{n\gamma }$ is an integer. Since the gauge function $\chi (%
\vec{x})$ is just a function in 3-dimensional space it is the same for all
particles and the integral on the left is independent of $n$. This implies
that the charge $q_{n}$ divided by an integer $\Delta k_{n\gamma }$ must be
independent of $n$ which means that $q_{n}$ must be an integer multiple of
some basic charge $q_{0}$. We conclude that the charges $q_{n}$ are
quantized.

The issue of charge quantization is ultimately the issue of deciding which
is the gauge group that generates electromagnetic interactions. We could for
example decide to restrict the gauge transformations to single-valued gauge
functions $\chi (\vec{x})$ so that (\ref{circ b}) is trivially satisfied
irrespective of the charges being quantized or not. Under such a restricted
symmetry group the single-valued (or double-valued) nature of the wave
function is unaffected by gauge transformations. If, on the other hand, the
gauge functions $\chi (\vec{x})$ are allowed to be multi-valued, then the
compatibility of the gauge transformation (\ref{GT 1}-\ref{GT 2}) with the
superposition principle demands that charges be quantized.

The argument above cannot fix the value of the basic charge $q_{0}$ because
it depends on the units chosen for the vector potential $A_{a}$. Indeed
since the dynamical equations show $q_{n}$ and $A_{a}$ appearing only in the
combination $q_{n}A_{a}$ we can change units by rescaling charges and
potentials according to $Cq_{n}=q_{n}^{\prime }$ and $A_{a}/C=A_{a}^{\prime
} $ so that $q_{n}A_{a}=q_{n}^{\prime }A_{a}^{\prime }~$. For conventional
units such that the basic charge is $q_{0}=e/3$ with $\alpha =e^{2}/\hbar
c=1/137$ the scaling factor is $C=(\alpha \hbar c)^{1/2}/3q_{0}$. A more
natural set of units might be to set $q_{0}=\hbar c$ so that all $\beta _{n}$%
s are integers and the gauge functions $\chi (\vec{x})$ are angles.\emph{\ }

A similar conclusion --- that charge quantization is a reflection of the
compactness of the gauge group --- can be reached following an argument due
to C. N. Yang \cite{Yang 1970}. Yang's argument assumes that a Hilbert space
has been established and one has access to the unitary representations of
symmetry groups. Yang considers a gauge transformation 
\begin{equation}
\Psi (x)\rightarrow \Psi (x)\exp i\tsum\nolimits_{n}\frac{q_{n}}{c}\chi (%
\vec{x}_{n})~,  \label{GT 3}
\end{equation}%
with $\chi (\vec{x})$ independent of $\vec{x}$. If the $q_{n}$s are not
commensurate there is no value of $\chi $ (except $0$) that makes (\ref{GT 3}%
) be the identity transformation. The gauge group --- translations on the
real line --- would not be compact. If, on the other hand, the charges are
integer multiples of a basic charge $q_{0}$, then two values of $\chi $ that
differ by an integer multiple of $2\pi c/q_{0}$ give identical
transformations and the gauge group is compact. In the present ED
derivation, however, we deal with the space $T^{\ast }\mathbf{P}$ which is a
complex projective space. We cannot adopt Yang's argument because a gauge
transformation $\chi $ independent of $\vec{x}$ is already an identity
transformation --- it leads to an equivalent state in the same ray --- and
cannot therefore lead to any constraints on the allowed charges.

\section{Classical limits and the Bohmian limit}

\subsection{Classical limits}

There are two classical limits that one might wish to consider. One is the
mathematical limit $\hbar \rightarrow 0$. Taking $\hbar \rightarrow 0$
leaves unchanged both the velocities $v_{n}^{a}$ of the particles, eq.(\ref%
{drift velocity}), and the probability flow, eq.(\ref{CE c}). The main
effect is to suppress the quantum potential so that eq.(\ref{HJ}) becomes
the classical Hamilton-Jacobi equation. The symplectic form, eq.(\ref{sympl
form c}), survives unscathed but the metric and the complex structures, eqs.(%
\ref{ePS c}) and (\ref{J tensor}), do not. But this is not quite classical
mechanics. Since the velocity fluctuations, eq.(\ref{fluct c}), remain
unaffected the resulting dynamics is a non-dissipative version of the
classical Oernstein-Uhlenbeck Brownian motion. To recover a deterministic
classical mechanics one must also take the limit $\eta \rightarrow 0$ .

The other classical limit arises in the more physically relevant situation
where one deals with a system with a large number $N$ of particles --- for
example, a macromolecule or a speck of dust --- and one wishes to study the
motion of an effective macrovariable such as the center of mass (CM), eq.(%
\ref{CM}). The large $N$ limit of ED with particles undergoing an ES
Brownian motion was studied in \cite{Demme Caticha 2016}. The same argument
goes through essentially unchanged for the OU Brownian motion discussed
here. Skipping all details we find that as a consequence of the central
limit theorem the continuity equation for $\rho _{\text{cm}}(X^{a})$\ and
the velocity fluctuations are given by the analogues of (\ref{CE}) and (\ref%
{fluct c}) for a single particle of mass $M=\tsum\nolimits_{n=1}^{N}m_{n}$,%
\begin{equation}
\partial _{t}\rho _{\text{cm}}=\frac{\partial }{\partial X^{a}}\left( \rho _{%
\text{cm}}V^{a}\right) \quad \text{with}\quad V^{a}=\frac{\langle \Delta
X^{a}\rangle }{\Delta t}=\frac{1}{M}\frac{\partial \Phi _{\text{cm}}}{%
\partial X^{a}}~,  \label{CE cm}
\end{equation}%
\begin{equation}
\left\langle \left( \frac{\Delta X^{a}}{\Delta t}-V^{a}\right) \left( \frac{%
\Delta X^{b}}{\Delta t}-V^{b}\right) \right\rangle =\frac{\eta \Delta t}{M}.
\label{fluct cm}
\end{equation}%
We also find that under rather general conditions the CM\ motion decouples
from the motion of the component particles and obeys the single particle HJ
equation 
\begin{equation}
-\partial _{t}\Phi _{\text{cm}}=\frac{1}{2M}\left( \frac{\partial \Phi _{%
\text{cm}}}{\partial X^{a}}\right) ^{2}-\frac{\hbar ^{2}}{2M}\frac{\nabla
^{2}\rho _{\text{cm}}^{1/2}}{\rho _{\text{cm}}^{1/2}}+V_{\text{ext}}(X)~.
\label{CM HJ}
\end{equation}%
In the large $N$ limit $M\sim O(N)$ and we obtain a finite velocity $V^{a}$
in (\ref{CE cm}) provided $\Phi _{\text{cm}}\sim O(N)$. In eq.(\ref{CM HJ})
we see that for a sufficiently large system the quantum potential for the
CM\ motion vanishes. Therefore, for $N\rightarrow \infty $, the CM follows
smooth trajectories described by a classical Hamilton-Jacobi equation.
Furthermore, eq.(\ref{fluct cm}) shows that as $N\rightarrow \infty $ the
velocity fluctuations vanish irrespective of the value of $\eta $. This is a
truly deterministic classical mechanics.

An important feature of this derivation is that $\hbar $ and $\eta $ remain
finite which means that a mesoscopic or macroscopic object will behave
classically while all its component particles remain fully quantum
mechanical.

\subsection{The Bohmian limit}

ED models with different values of $\eta $ lead to the same Schr\"{o}dinger
equation. In other words, different sub-quantum models lead to the same
emergent quantum behavior. The limit of vanishing $\eta $ deserves
particular attention because the velocity fluctuations, eq.(\ref{fluct c}),
are suppressed and the motion becomes deterministic. This means that ED
includes a Bohmian form of quantum mechanics \cite{Bohm 1952}-\cite{Holland
1993} as a special limiting case --- but with the important caveat that the
difference in physical interpretation remains enormous. It is only with
respect to the mathematical formalism that ED includes Bohmian mechanics as
a special case.

Bohmian mechanics attempts to provide an actual description of reality. In
the Bohmian view the universe consists of real particles that have definite
positions and their trajectories are guided by a real field, the wave
function $\Psi $. Not only does this pilot wave live in $3N$-dimensional
configuration space but it manages to act on the particles without the
particles reacting back upon it. These are peculiarities that have stood in
the way of a wider acceptance of the Bohmian interpretation. In contrast,
ED's pragmatic goal is much less ambitious: to make the best possible
predictions on the basis of very incomplete information. As in Bohmian
mechanics, in ED the particles have definite positions and its formalism
includes a function $\Phi $ that plays the role of a pilot wave. But $\Phi $
is an epistemic tool for reasoning; it is not meant to represent anything
real. There is no implication that the particles move the way they do
because they are pushed around by a pilot wave or by some stochastic force.
In fact ED is silent on the issue of what if anything is pushing the
particles. What the probability $\rho $ and the phase $\Phi $ are designed
to do is not to guide the particles but to guide our inferences. They guide
our expectations of where and when to find the particles but they do not
exert any causal influence on the particles themselves.

\section{Hilbert space}

The formulation of the ED of spinless particles is now complete. We note, in
particular, that the notion of Hilbert spaces turned out to be unnecessary
to the formulation of quantum mechanics. As we shall see next, while
strictly unnecessary in principle, the introduction of Hilbert spaces is
nevertheless very convenient for calculational purposes.

\paragraph*{A vector space ---}

As we saw above the infinite-dimensional e-phase space --- the cotangent
bundle$\ T^{\ast }\mathbf{P}$ --- is difficult to handle. The problem is
that the natural coordinates are probabilities $\rho _{x}$ which, due to the
normalization constraint, are not independent. In a discrete space one could
single out one of the coordinates and its conjugate momentum and then
proceed to remove them. Unfortunately, with a continuum of coordinates and
momenta the removal is not feasible. The solution is to embed $T^{\ast }%
\mathbf{P}$ in a larger space $T^{\ast }\mathbf{P}^{+1}$. This move allows
us to keep the natural coordinates $\rho _{x}$ but there is a price: we are
forced to deal with a constrained system and its attendant gauge symmetry.

We\ also saw that the geometry of the embedding space was not fully
determined: any spherically symmetric space would serve our purposes. This
is a freedom we can further exploit. For calculational purposes the
linearity of the Schr\"{o}dinger equation (\ref{sch b}) is very convenient
but its usefulness is severely limited by the normalization constraint. If $%
\Psi _{1}$ and $\Psi _{2}$ are flows in $T^{\ast }\mathbf{P}$ then the
superposition $\Psi _{3}$ in (\ref{psi3}) will also be a flow in $T^{\ast }%
\mathbf{P}$ but only if the coefficients $\alpha _{1}$ and $\alpha _{2}$ are
such that $\Psi _{3}$ is properly normalized. This restriction can be
removed by choosing the extended embedding space $T^{\ast }\mathbf{P}^{+1}$
to be flat --- just set $A=0$ and $B=1$ in eq.(\ref{TP+1 metric}). (The fact
that this space is flat is evident in the metric (\ref{spherical metric})
for the discrete case.) We emphasize that this choice is not at all
obligatory; it is optional.

The fact that in the flat space $T^{\ast }\mathbf{P}^{+1}$ superpositions
are allowed for arbitrary constants $\alpha _{1}$ and $\alpha _{2}$ means
that $T^{\ast }\mathbf{P}^{+1}$ is not just a manifold; it is also a vector
space. Each point $\Psi $ in $T^{\ast }\mathbf{P}^{+1}$ is itself a vector.
Furthermore, since the vector tangent to a curve is just a difference of two
vectors\ $\Psi $ we see that that points on the manifold and vectors tangent
to the manifold are objects of the same kind. In other words, the tangent
spaces $T[T^{\ast }\mathbf{P}^{+1}]_{\Psi }$ are identical to the space $%
T^{\ast }\mathbf{P}^{+1}$ itself.

The symplectic form $\Omega $ and the metric tensor $G$ on the extended
space $T^{\ast }\mathbf{P}^{+1}$ are given by eq.(\ref{sympl form d}) and (%
\ref{metric Psi b}). Since they are tensors $\Omega $ and $G$ are meant to
act on vectors but now they can also act on all points $\Psi \in T^{\ast }%
\mathbf{P}^{+1}$ and not just on those that happen to be normalized and
gauge fixed according to (\ref{TGF}). For example, the action of the mixed
tensor $J$, eq.(\ref{complex structure J}), on a wave function $\Psi $ is 
\begin{equation}
J^{\mu x}{}_{\nu x^{\prime }}\Psi ^{\nu x^{\prime }}=%
\begin{bmatrix}
i & 0 \\ 
0 & -i%
\end{bmatrix}%
\binom{\Psi _{x}}{i\hbar \Psi _{x}^{\ast }}=\binom{i\Psi _{x}}{i\hbar (i\Psi
_{x})^{\ast }}~,
\end{equation}%
which indicates that $J$ plays the role of multiplication by $i$, that is,
when acting on a point $\Psi $ the action of $J$ is $\Psi \overset{J}{%
\rightarrow }i\Psi $.

\paragraph*{Dirac notation ---}

We can at this point introduce the Dirac notation to represent the wave
functions $\Psi _{x}$ as vectors $|\Psi \rangle $ in a Hilbert space. The
scalar product $\langle \Psi _{1}|\Psi _{2}\rangle $ is defined using the
metric $G$ and the symplectic form $\Omega $, 
\begin{equation}
\langle \Psi _{1}|\Psi _{2}\rangle \overset{\text{def}}{=}\frac{1}{2\hbar }%
\int dx\,dx^{\prime }\left( \Psi _{1x},i\hbar \Psi _{1x}^{\ast }\right)
\left( G+i\Omega \right) \binom{\Psi _{2x^{\prime }}}{i\hbar \Psi
_{2x^{\prime }}^{\ast }}~.
\end{equation}%
A straightforward calculation gives 
\begin{equation}
\langle \Psi _{1}|\Psi _{2}\rangle =\int dx\,\Psi _{1}^{\ast }\Psi _{2}~.
\end{equation}%
The map $\Psi _{x}\leftrightarrow |\Psi \rangle $ is defined by 
\begin{equation}
|\Psi \rangle =\int dx|x\rangle \Psi _{x}\quad \text{where}\quad \Psi
_{x}=\langle x|\Psi \rangle ~,
\end{equation}%
where, in this \textquotedblleft position\textquotedblright\
representation,\ the vectors $\{|x\rangle \}$ form a basis that is
orthogonal and complete, 
\begin{equation}
\langle x|x^{\prime }\rangle =\delta _{xx^{\prime }}\quad \text{and}\quad
\int dx\,|x\rangle \langle x|~=\hat{1}~.
\end{equation}

\paragraph*{Hermitian and unitary operators ---}

The bilinear Hamilton functionals $\tilde{Q}[\Psi ,\Psi ^{\ast }]$ with
kernel $\hat{Q}(x,x^{\prime })$ in eq.(\ref{bilinear hamiltonian}) can now
be written in terms of a Hermitian operator $\hat{Q}$ and its matrix
elements, 
\begin{equation}
\tilde{Q}[\Psi ,\Psi ^{\ast }]=\langle \Psi |\hat{Q}|\Psi \rangle \quad 
\text{and}\quad \hat{Q}(x,x^{\prime })=\langle x|\hat{Q}|x^{\prime }\rangle
~.
\end{equation}%
The corresponding Hamilton-Killing flows are given by 
\begin{equation}
i\hbar \frac{d}{d\lambda }\langle x|\Psi \rangle =\langle x|\hat{Q}|\Psi
\rangle \quad \text{or}\quad i\hbar \frac{d}{d\lambda }|\Psi \rangle =\hat{Q}%
|\Psi \rangle ~.
\end{equation}%
These flows are described by unitary transformations 
\begin{equation}
|\Psi (\lambda )\rangle =\hat{U}_{Q}(\lambda )|\Psi (0)\rangle \quad \text{%
where}\quad \hat{U}_{Q}(\lambda )=\exp \left( -\frac{i}{\hbar }\hat{Q}%
\lambda \right) ~.
\end{equation}

\paragraph*{Commutators ---}

The Poisson bracket of two Hamiltonian functionals $\tilde{U}[\Psi ,\Psi
^{\ast }]$ and $\tilde{V}[\Psi ,\Psi ^{\ast }]$,

\begin{equation*}
\{\tilde{U},\tilde{V}\}=\int dx\left( \frac{\delta \tilde{U}}{\delta \Psi
_{x}}\frac{\delta \tilde{V}}{\delta i\hbar \Psi _{x}^{\ast }}-\frac{\delta 
\tilde{U}}{\delta i\hbar \Psi _{x}^{\ast }}\frac{\delta \tilde{V}}{\delta
\Psi _{x}}\right) ~,
\end{equation*}%
can be written in terms of the commutator of the associated operators,then 
\begin{equation}
\{\tilde{U},\tilde{V}\}=\frac{1}{i\hbar }\langle \Psi |[\hat{U},\hat{V}%
]|\Psi \rangle ~.
\end{equation}%
Thus the Poisson bracket is the expectation of the commutator. This \emph{%
identity} is much sharper than Dirac's pioneering discovery that the quantum
commutator of two $q$-variables is \emph{analogous} to the Poisson bracket
of the corresponding classical variables. Further parallels between the
geometric and the Hilbert space formulation of QM can be found in \cite%
{Kibble 1979}-\cite{Elze 2012}.

\section{Remarks on ED and Quantum Bayesianism}

\label{final remarks} Having discussed the ED approach in some detail it is
now appropriate to comment on how ED differs from the interpretations known
as Quantum Bayesianism \cite{Brun et al 2001}-\cite{Caves et al 2002b} and
its closely related descendant QBism \cite{Fuchs Schack 2013}\cite{Fuchs et
al 2014}; for simplicity, I shall refer to both as QB. Both ED and QB adopt
an epistemic degree-of-belief concept of probability but there are important
differences:

\textbf{(a)} QB adopts a personalistic de Finetti type of Bayesian
interpretation while ED adopts an impersonal entropic Bayesian
interpretation somewhat closer but not identical to Jaynes' \cite{Jaynes
1983}-\cite{Caticha 2014}. In ED the probabilities do not reflect the
subjective beliefs of any particular person. They are tools designed to
assist us in those all too common situations in which are confused and due
to insufficient information we do not know what to believe. The
probabilities will then provide guidance as to what agents ought to believe
if only they were ideally rational. More explicitly, probabilities in ED
describe the objective degrees of belief of ideally rational agents who have
been supplied with the maximal allowed information about a particular
quantum system.

\textbf{(b)} ED derives or reconstructs the mathematical framework of QM ---
it explains where the symplectic, metric, and complex structures, including
Hilbert spaces and time evolution come from. In contrast, at its current
stage of development QB consists of appending a Bayesian interpretation to
an already existing mathematical framework. Indeed, assumptions and concepts
from quantum information are central to QB and are implicitly adopted from
the start. For example, a major QB concern is the justification of the Born
rule starting from the Hilbert space framework while ED starts from
probabilities and its goal is to justify the construction of wave functions;
the Born rule follows as a trivial consequence.

\textbf{(c)} ED is an application of entropic/Bayesian inference. Of course,
the choices of variables and of the constraints that happen to be physically
relevant are specific to our subject matter --- quantum mechanics --- but
the inference method itself is of universal applicability. It applies to
electrons just as well as to the stock market or to medical trials. In
contrast, in QB the personalistic Bayesian framework is not of universal
validity. For those special systems that we call `quantum' the inference
framework is itself modified into a new \textquotedblleft Quantum-Bayesian
coherence\textquotedblright\ in which the standard Bayesian inference must
be supplemented with concepts from quantum information theory. The
additional technical ingredient is a hypothetical structure called a
\textquotedblleft symmetric informationally complete
positive-operator-valued measure\textquotedblright . In short, in QB Born's
Rule is not derived but constitutes an addition beyond the raw probability
theory.

\textbf{(d)} QB is an anti-realist neo-Copenhagen interpretation; it accepts
complementarity. (Here complementarity is taken to be the common thread that
runs through all Copenhagen interpretations.) Probabilities in QB refer to
the outcomes of experiments and not to ontic pre-existing values. In
contrast, in ED probabilities refer to ontic positions --- including the
ontic positions of pointer variables. In the end, this is what solves the
problem of quantum measurement (see \cite{Johnson Caticha 2011}\cite%
{Vanslette Caticha 2016}).

\section{Some final remarks}

We conclude with a summary of the main assumptions:

\begin{description}
\item 
\begin{itemize}
\item \qquad Particles have definite but unknown positions and follow
continuous trajectories.
\end{itemize}

\item 
\begin{itemize}
\item \qquad The probability of a short step is given by the method of
maximum entropy subject to a drift potential constraint that introduces
directionality and correlations, plus gauge constraints that account for
external electromagnetic fields.
\end{itemize}

\item 
\begin{itemize}
\item \qquad The accumulation of short steps requires a notion of time as a
book-keeping device. This involves the introduction of the concept of an
instant and a convenient definition of the duration between successive
instants.
\end{itemize}

\item 
\begin{itemize}
\item \qquad The e-phase space $\{\rho ,\Phi \}$ has a natural symplectic
geometry that results from treating the pair $(\rho _{x},\Phi _{x})$ as
canonically conjugate variables.
\end{itemize}

\item 
\begin{itemize}
\item \qquad The information geometry of the space of probabilities is
extended to the full e-phase space by imposing the latter be spherically
symmetric.
\end{itemize}

\item 
\begin{itemize}
\item \qquad The drift potential constraint is updated instant by instant in
such a way as to preserve both the symplectic and metric geometries of the
e-phase space.
\end{itemize}
\end{description}

\noindent The resulting entropic dynamics is described by the Schr\"{o}%
dinger equation. Different sub-quantum Brownian motions all lead to the same
emergent quantum mechanics. In previous work we dealt with an
Einstein-Smoluchowski process; here we have explored an Oernstein-Uhlenbeck
process. Other \textquotedblleft fractional\textquotedblright\ Brownian
motions might be possible but have not yet been studied.

A natural question is whether these different sub-quantum Brownian motions
might have observable consequences. At this point our answer can only be
tentative. To the extent that we have succeeded in deriving QM and not some
other theory one should not expect deviations in the predictions for those
standard experiments that are the subject of the standard quantum theory ---
at least not in the nonrelativistic and microscopic regimes. As the ED
program is extended to other regimes involving higher energies and/or
gravity it is quite possible that those different sub-quantum motions might
not be empirically equivalent.

ED achieves ontological clarity by sharply separating the ontic elements
from the epistemic elements --- positions of particles on one side and
probabilities $\rho $ and phases $\Phi $ on the other. ED is a dynamics of
probabilities and not a dynamics of particles. Of course, if probabilities
at one instant are large in one place and at a later time they are large in
some other place one infers that the particles must have moved --- but
nothing in ED describes what it is that has pushed the particles around. ED
is a mechanics without a mechanism.

We can elaborate on this point from a different direction. The empirical
success of ED suggests that its epistemic probabilities are in agreement
with ontic features of the physical world. It is highly desirable to clarify
the precise nature of this agreement. Consider, for example, a fair die. Its
property of being a perfect cube is an ontic property of the die which is
reflected at the epistemic level in the equal\ assignment of probabilities
to each face of the die. In this example we see that the epistemic
probabilities achieve objectivity, and therefore usefulness, by
corresponding to something ontic. The situation in ED is similar except for
one crucial aspect. The ED probabilities are objective and they are\
empirically successful. They must therefore reflect something real. However,
it is not yet known what those underlying ontic properties might possibly
be. Fortunately, for the purposes of making predictions knowing those
epistemic probabilities is all we need.

The trick of embedding the e-phase space $T^{\ast }\mathbf{P}$ in a \emph{%
flat} vector space $T^{\ast }\mathbf{P}^{+1}$ is clever but optional. It
allows one to make use of the calculational advantages of linearity. This
recognition that Hilbert spaces are not fundamental is one of the
significant contributions of the entropic approach to our understanding of
QM. The distinction --- whether Hilbert spaces are necessary in principle as
opposed to merely convenient in practice --- is not of purely academic
interest. It can be important in the search for a quantum theory that
includes gravity: Shall we follow the usual approaches to quantization that
proceed by replacing classical dynamical variables by an algebra of linear
operators acting on some abstract space? Or, in the spirit of an entropic
dynamics, shall we search for an appropriately constrained dynamics of
probabilities and information geometries? First steps towards formulating a
first-principles theory along this lines are given in \cite{Ipek Caticha
2019}\cite{Caticha 2019}.

\subparagraph*{Acknowledgments}

I would like to thank M. Abedi, D. Bartolomeo, C. Cafaro, N. Carrara, N.
Caticha, F. Costa, S. DiFranzo, K. Earle, A. Giffin, S. Ipek, D.T. Johnson,
K. Knuth, O. Lunin, S. Nawaz, P. Pessoa, M. Reginatto, C. Rodr\'{\i}guez,
and K. Vanslette, for valuable discussions on entropic inference and
entropic dynamics and for their many insights and contributions at various
stages of this program.

\end{document}